\documentclass{article} 

\usepackage{graphicx,color,amsmath,amsfonts,amssymb,times,url}
\begin{document}

\title{Order patterns, their variation and change points in financial time series and Brownian motion}  

\author{Christoph Bandt\\ 
University of Greifswald, Germany, 
bandt@uni-greifswald.de}
\date{\today}
\maketitle

\begin{abstract}
Order patterns and permutation entropy have become useful tools for studying biomedical, geophysical or climate time series.  Here we study day-to-day market data, and Brownian motion which is a good model for their order patterns.  A crucial point is that for small lags (1 up to 6 days), pattern frequencies in financial data remain essentially constant.  The two most important order parameters of a time series are turning rate and up-down balance. For change points in EEG brain data, turning rate is excellent while for financial data, up-down balance seems the best. The fit of Brownian motion with respect to these parameters is tested, providing a new version of a forgotten test by Bienaym\'e. 
\end{abstract}

\section{Overview} \label{over}
Given a finite series of data, time series analysis attempts to find laws of the mechanism or process which generated the data. One tool to this end are order patterns, defined as permutations in section \ref{face}. Their probabilities are properties of the random process which can be easily estimated from the data. The permutation entropy \cite{BP} measures the variety of patterns.  Applications to brain and heart data concern epilepsy \cite{Bru,Fe}, Alzheimer's disease \cite{Mor}, effect of anaesthesia \cite{OSD}, and  cardiac dynamics \cite{Par12,Chi,McCullough17}. For applications to physics, geophysics, environtmental and climate data see the surveys \cite{ZZRP,AKU} and the collection \cite{AKK}.  

Moreover, probabilities of specific order patterns can give important information. For chaotic dynamical systems, certain patterns will not appear \cite{Ami}. Different versions of permutation entropy and the number of forbidden patterns were used to quantify and monitor the efficiency of stock \cite{ZZTP} and bond markets \cite{ZB12,ZB16}. The probabilities of patterns can be combined to form correlation functions \cite{Ba15,Ba17}. The dependence of economical time series was studied in \cite{Schnurr} by means of order patterns. Statistical theory of order pattern estimators was developed recently by \cite{SD,BDS}, earlier work by \cite{BS,SK11} was restricted to Gaussian processes.

Several authors have studied change points and segmentation of time series by means of order patterns \cite{SKC,UK18}. The interesting approach in \cite{Schnurr,SD} uses bivariate data while we shall study the univariate case only.  A very impressive example is sleep stage classification using high-frequency data of a single EEG channel \cite{NG,KL,Ba17,Ba19}. Here we have an abundance of data and few statistical problems. In the next section we will briefly describe this application.

Our main theme are order patterns in day-to-day financial data and their statistics.  
The zigzag patterns of up and down have long been considered by stock traders. Today they are certainly incorporated in secret algorithms for high-frequency trade. We are interested in the statistical error of estimates of pattern probabilties, and in the change point problem. We introduce our standard example of oil prices in section \ref{pat4}.  

A curious feature of such examples is their apparent`self-similarity' studied by Mandelbrot since the 1960s \cite{ManF}. This property implies equality of frequencies of patterns for different lags, which can be compared with simulated samples of Brownian motion and other Levy processes. One goal of the paper is to develop tests which decide whether Brownian motion is an appropriate model. In section \ref{brow}  this is done by simulation, in section \ref{taube} rigorously. 

It seems that for a large part, `order self-similarity' is due to the uneven sampling of market data, with missing weekends and holidays. In contrast to measurements done in nature, the observed values are not varying in natural time. Their change is triggered by trade. Volume and number of buying and selling orders provide alternative scales to natural time.  Under such conditions, classical tools like autocorrelation become useless. It may be a necessity to postulate equality of pattern frequencies for small lags. Our discussion in section \ref{daily} shows that the assumption is justified and can be used to improve estimates of pattern frequencies.

Our expectations should be modest: with at most a few thousand data, statistical accuracy will not be magic.
In section \ref{taube} we introduce the two most stable and interpretable order parameters, up-down balance and turning rate.  They are used to provide statistical tests for comparing models and data series. Actually, a first test for the turning rate was suggested already in \cite{bien1,bien2}. The basic message is that pattern statistics has the $1/\sqrt{n}$ accuracy coming from binomial distribution. In Bienaym\`e's case it is even better because of negative correlations. 
In section \ref{chang} we use order patterns to determine change points in financial time series. It turns out that up-down balance is the most appropriate parameter.

\section{A big data series from medicine} \label{slee}
We start with a brief description of segmentation of big series of brain data by means of the order structure. This has been done by many authors, cf. \cite{OSD,ODRL,NG,KL,SKC,Ba17,UK18}. Our version in \cite{Ba18} is particularly simple. We considered electroencephalographic (EEG) sleep data measured by \cite{Te}, and publicly available at pysionet \cite{physio}.
Healthy volunteers and patients were measured for one night in hospital with a sample rate of 500 Hz. These are multivariate data, as is the standard in sleep labs, and sleep stages (wake - REM sleep - sleep stages 1 to 4) were annotated by trained experts for every epoch of 30 seconds.  For our classification, we used only two EEG channels, each with 20 million data points for one night. 

\begin{figure}[h]  
\centering
\includegraphics[width=.8\textwidth]{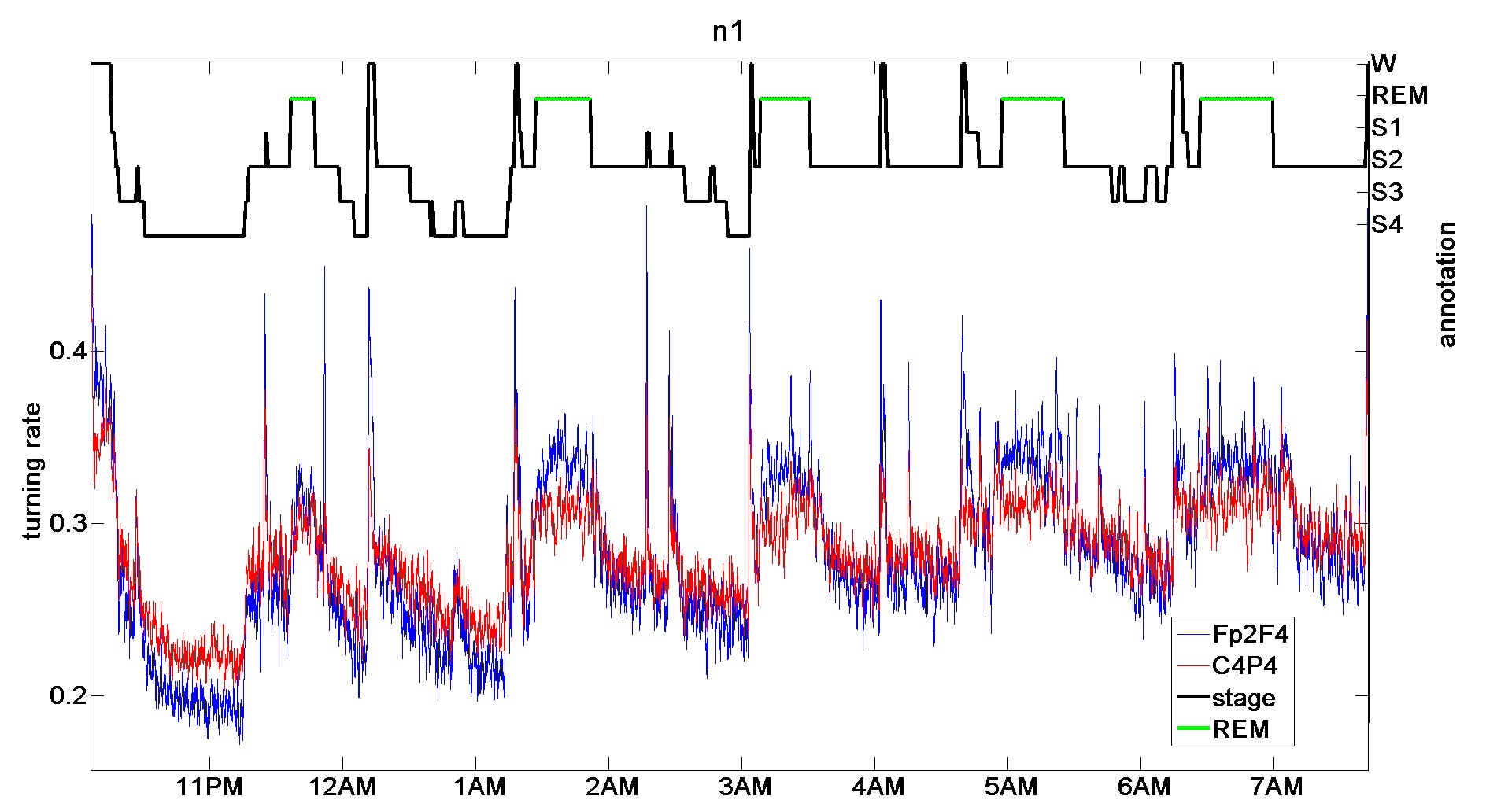}
\caption{Sleep stage classification by turning rate} \label{sl1}
\end{figure}

We determined a score for each epoch in an extremely simple way: we just counted the relative number of local maxima and minima in the time series. In Figure \ref{sl1} it can be seen that the result almost coincides with the expert's annotation. In white noise, the turning rate - the relative amount of local extrema - is known to be $\frac{2}{3}$ \cite{bien1}.   In Figure \ref{sl1} we see turning rates between 0.35 and 0.45 when the proband is awake, with larger rates in the more frontal channel. In REM sleep, rates were between 0.3 and 0.35, also larger for the more frontal channel.  When the patient was in non-REM sleep, turning rates went below 0.3, with larger values in the central channel. The smallest turning rate is below 0.2, in the first phase of deep sleep. After midnight, turning rates of sleep stages 3 and 4 gradually increased. 
Such results, with individual differences, were observed for about 20 persons measured by \cite{Te} with 500 Hz while for lower sampling rate, and for other data contaminated by 50 Hz noise from the power net, the results were less impressive \cite{Ba18}. 
  
Here we looked for many change points, and a very simple method was sufficient. The main thing was to express the structure by the appropriate parameter  (see section \ref{taube} for details). Of course, this is preliminary research, and much work has to be done to make it an applicable method.  But we are in a comfortable position with 15000 data for each epoch.
For this application, a precision of $\pm 10$ seconds for change points is fine, so an error of $\pm 1000$ time points does not count. The data contain plenty of artefacts when the patient moves, rolls with the eyes in dream, gnabs with the teeth etc. A few bad epochs can be tolerated, however. Most of them are annotated `wake' in Figure \ref{sl1}. The sliding window technique works well. We just determine turning rate for epochs of 30 seconds. For greater detail, we can take overlapping windows, with a shift of one second only \cite{Ba18}. There are no serious statistical problems.

In this paper on historic economical data, the situation is quite different. We have a few thousand values altogether. We must think about which structural parameters can be estimated with reasonable accuracy, and we have to deal with statistical errors.

\section{Pattern frequencies as one facet of a random process} \label{face}
As we said, we try to find properties of the process which generates the data. We begin with a brief glance at the theory.
A stationary random process $X_1,X_2,...$ in discrete time, ${\mathbb N}=\{ 1,2,...\}$ is essentially defined by its distribution, a probability measure $P$ on Borel sets $B$ of the space  $\Omega={\mathbb R}^{\mathbb N}$ of all possible infinite time series $y_1,y_2,...,$ which is invariant under time shifts. It is enough to know the probabilities of those Borel sets $B\subset {\mathbb R}^n$ which depend only on the first values $y_1,...,y_n,$ for arbitrary $n.$ 

A statistician will ask how probabilities $P(B)$ are estimated from a finite time series $x=x_1,x_2,...,x_T .$ Since we assume stationarity, an estimate $\widehat{P(B)}$ is the relative frequency of those vectors  $(x_{t+1},...,x_{t+n})$  which belong to the set $B,$ for $t=0,...,T-n.$ As a formula,
\begin{equation}
\widehat{P(B)} = \frac{1}{T+1-n}\ \# \{ t\  |\  0\le t\le T-n, \ (x_{t+1},...,x_{t+n}) \in B \} \ .
\label{estib}\end{equation}

Here $\# A$ denotes the cardinality of a set $A.$ Let us assume that the stationary process is ergodic. Then a classical theorem of Birkhoff says that for $T\to\infty ,$ the estimate $\widehat{P(B)}$ will converge to the true value $P(B),$ with probability one. 

Alright! But in practice, $T$ is fixed and fairly small, say $T=1000.$   To get 
reasonable estimates, $B$ should have a simple structure, like $\{x |  x_1< b\}$ for given numbers $b,$ which leads to the one-dimensional distribution.  Multidimensional Borel sets have rarely been studied. Usually one prefers to study autocorrelation,  based on averages over $t$ of functions $x_{t+1}x_{t+n}$ for $n=1,2,... .$ 

\begin{figure}[h]  
\centering
\includegraphics[width=.8\textwidth]{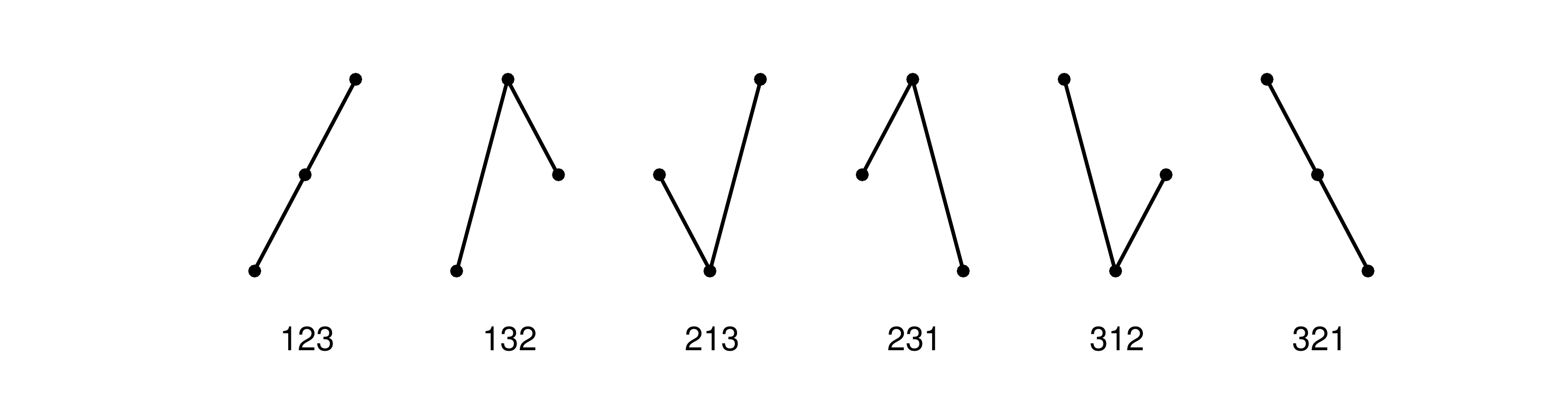}
\caption{The six order patterns of length 3} \label{ord3}
\end{figure}

There is, however, one class of multidimensional sets $B$ for which estimation makes sense: the sets $\{x |  x_j< x_k\}$ and their intersections. The set $\{x |  x_1< x_3 \}$ describes a halfplane in the $x_1x_3$-plane, and  $B=\{x |  x_1< x_2< x_3\}$ describes a `cone', both in ${\mathbb R}^3$ and $\Omega .$ The latter set $B$ is symbolized as the first order pattern in Figure \ref{ord3}. There are five other patterns or permutations $\pi .$ For example, $\pi=231$ corresponds to $B=\{x |  x_3< x_1< x_2\} .$

Note that a permutation $\pi$ of length $n$ is a one-to-one mapping from $\{ 1,...,n\}$ onto itself. Figure \ref{ord3} shows the graphs of the six permutations of length 3. We shall say that $n$ equally spaced values  $x_{t+d},x_{t+2d},..., x_{t+nd}$ in our time series represent the permutation $\pi$ of length $n$ if they are in the same order. That is, $\pi (j)<\pi (k)$ if and only if  $x_{t+jd}<x_{t+kd},$ for $1\le j< k\le n.$ Now we can count relative frequencies of permutations in the time series, which we shall denote  $p_\pi (d).$  Let
\begin{equation}
p_\pi (1) = \frac{1}{T+1-n}\ \# \{ t\  |\  0\le t\le T-n, \ (x_{t+1},...,x_{t+n}) \mbox{ realizes } \pi \} \ .
\label{estip}\end{equation}

\begin{figure}[h]
\parbox{.44\textwidth}{
\includegraphics[width=.45\textwidth]{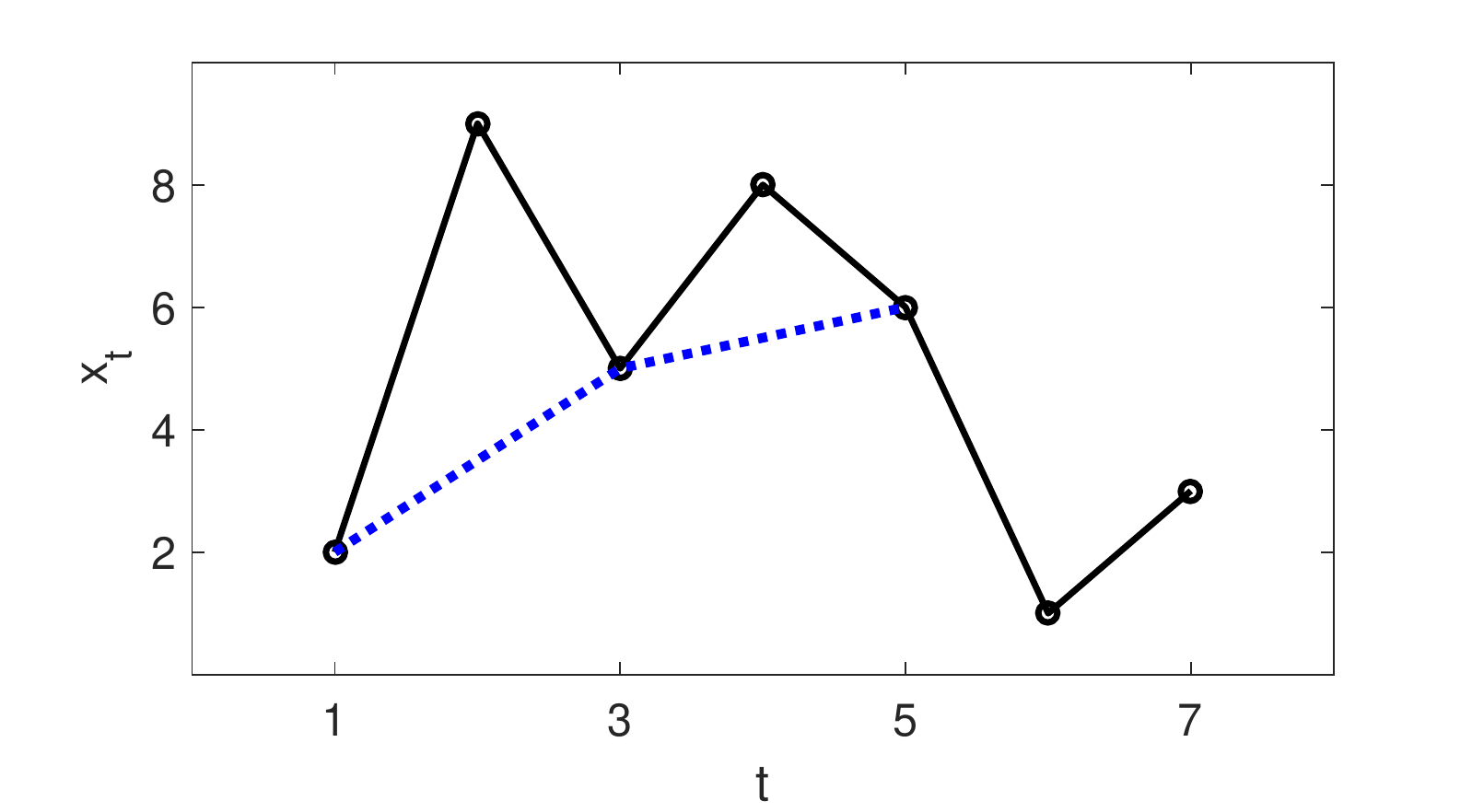}}\quad 
\parbox{.45\textwidth}{
\begin{tabular}{c|cccccc}
$\ \pi$&123&132&213&231&312&321\\ \hline
$d=1$&-&2&-&-&2&1\\
$d=2$&1&-&-&1&-&1\\ 
$d=3$&-&1&-&-&-&-\\ \hline
\end{tabular}}
\caption{Example time series and order pattern frequencies. The dotted line indicates $d=2.$} 
\label{fig2} \end{figure}

The case $d=1$ corresponds to successive values in the series, but sometimes it is good to compare values which are 2 or $d$ steps apart.  The formula \eqref{estib} for sets $B$ determined by a permutation $\pi$ of length $n$ and a lag $d\ge 1$ reads as follows.
\begin{equation}
p_\pi (d) = \frac{1}{T+d-nd}\ \# \{ t\  |\  -d+1\le t\le T-nd, \ (x_{t+d},...,x_{t+nd}) \mbox{ realizes } \pi \} \ .\tag{2d}
\label{estipd}\end{equation}
Only formally, we allow $t$ to be negative. Otherwise we had to consider permutations on $0,1,...,n-1$ as in \cite{AKU}. Figure \ref{fig2}, taken from \cite{Ba17}, 
shows a small example which leads to the frequencies  $p_{132}(1)=\frac25,\  p_{321}(2)=\frac13,$ and $p_{321}(3)=0.$

We have to note that our method excludes equality of values $x_j=x_k.$ In theory, equality happens only with probability zero when the one-dimensional distribution of the underlying process has a density function. In practice, equality of values will happen, but will be rare when the values have an accuracy of 5 or more decimals. As explained in \cite{Ba19}, 
pairs of equal values, as well as missing values, can just be disregarded in the calculation. A simpler practical method to eliminate equal values is to add a tiny white noise (uniform random numbers times $10^{-7},$ say) to the series $x.$
 
There are at least three good arguments for considering frequencies of such order patterns.  First, since they describe cones in $\Omega ,$ their estimates are robust with respect to noise and outliers, and not changed by an increasing nonlinear transformation of the data.  Next, order patterns can be evaluated very easily and extremely fast by computer. There are few possible sources of error. 

The most important argument for the use of order patterns are the
weak stationarity assumptions which are required. The process $X_1,X_2,...$ need not be stationary. It need only have stationary increments in the sense that for all $t,n\ge 1$
\begin{equation}
(X_{t+1}-X_t,...   ,X_{t+n}-X_t)\ \mbox{ has the same distribution as }\ (X_1,....,X_n)\ .
\label{incre}\end{equation}
For pairwise different numbers $x_t,...,x_{t+n},$ the vector $(x_{t+1}-x_t,...   ,x_{t+n}-x_t)$ always represents the same permutation as the vector $(x_{t+1},...,x_{t+n}).$ The same argument works for arbitrary $d.$ We obtain the following conclusion.\vspace{2ex}

{\bf Proposition 1. } \emph{ For an ergodic process $X_1,X_2,...$ with stationary increments, $p_\pi(d)$ in \eqref{estip} and  \eqref{estipd} is a consistent and unbiased estimator of the probability that the vector $(X_{t+d},...,X_{t+nd})$ represent the permutation $\pi ,$ for any fixed $t.$} \vspace{2ex}

This is a special case of the estimator \eqref{estib} which is known to be unbiased. 
The proposition says that our naive formula  \eqref{estipd} makes sense. In the sequel, we consider only patterns of length $n\le 4.$ 

Financial data have a random walk structure and are typical examples of processes with stationary increments. Brownian motion is their classical model. It is not stationary, and its autocorrelation function depends on two arguments. Its pattern frequencies as functions of the lag $d$ can be analytically determined, however \cite{BS}. \\

\section{Patterns of length four in a series of oil prices} \label{pat4}
Permutations of length 2 and 3 lead to two important parameters $\beta$ and $\tau ,$ as discussed in Section \ref{taube} below. To differentiate between various time series, we can use all 24 permutations of length 4. They provide detail information, and can be reasonably estimated when more than 1000 values are available. For a study of the 120 permutations of length 5 we would need much longer series.

For convenience, we shall assign  to each permutation a number between 1 and 24, according to lexicographic order, as indicated in Table \ref{lexi}.

\begin{table}[h]
\centering
\begin{tabular}{|c|c|c|c|c|c|c|c|c|c|c|c|} \hline
1&2&3&4&5&6&7&8&9&10&11&12\\
1234&1243&1324&1342&1423&1432&2134&2143&2314&2341&2413&2431\\ \hline 
13&14&15&16&17&18&19&20&21&22&23&24\\
3124&3142&3214&3241&3412&3421&4123&4132&4213&4231&4312&4321\\ \hline
\end{tabular}
\caption{Numbering of permutations of length 4}\label{lexi}
\end{table}

\begin{figure}[h]
\includegraphics[width=.48\textwidth]{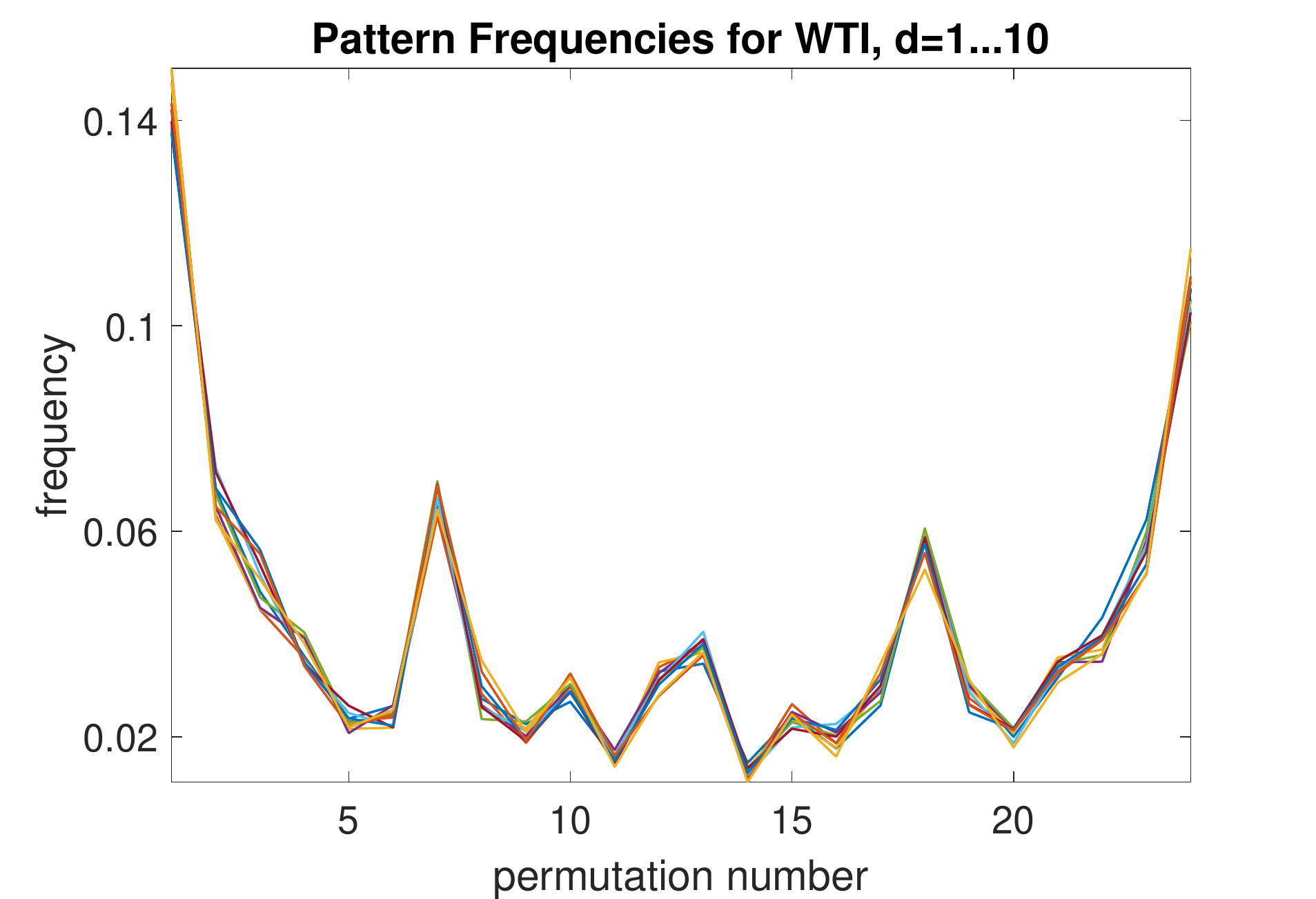} \quad 
\includegraphics[width=.48\textwidth]{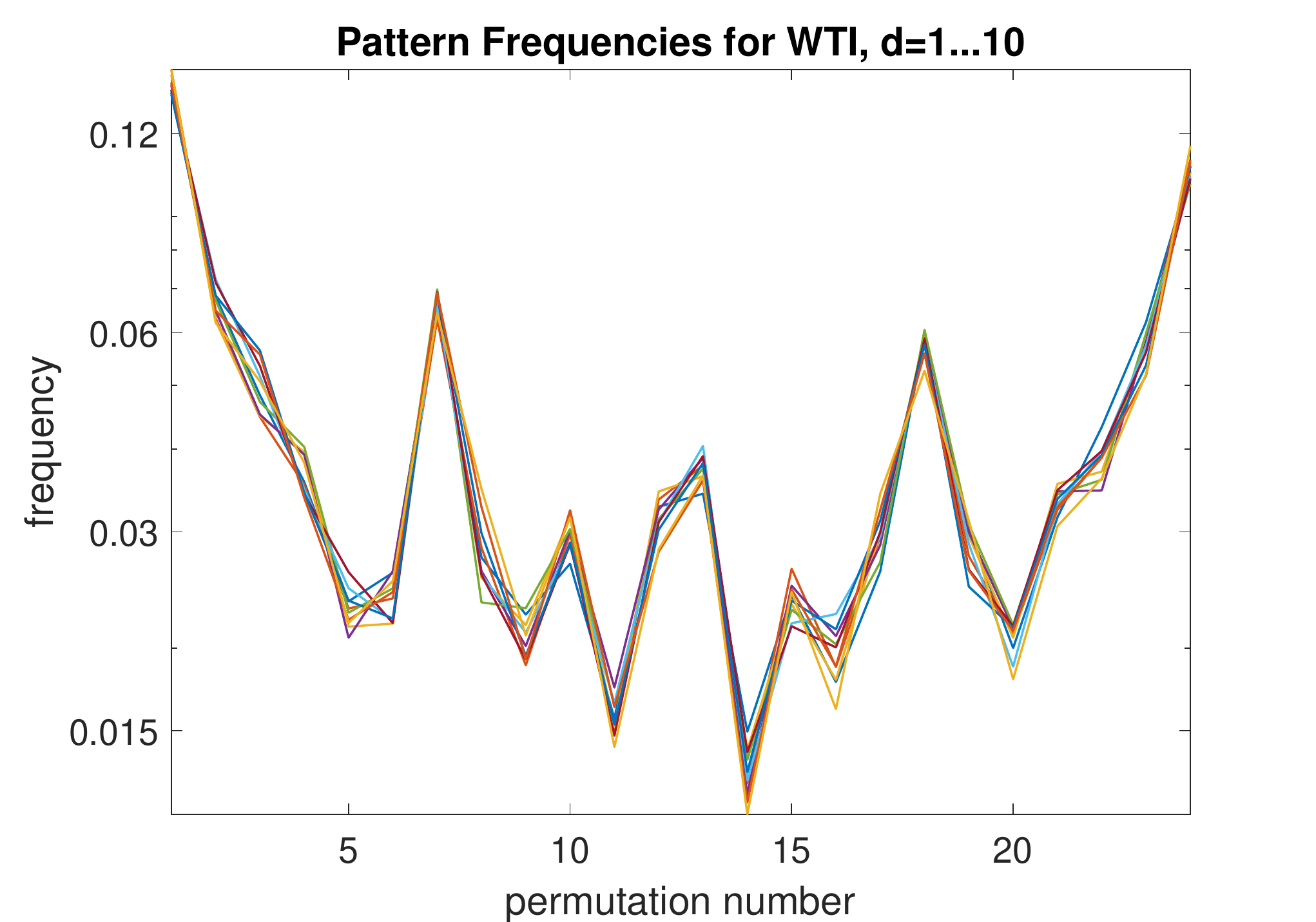}
\caption{Order four frequencies for the oil price series WTI, 1986-2019.} \label{WTI4}
\end{figure}

Figure \ref{WTI4} presents order four frequencies  (our abbreviation of `frequencies of order patterns of length 4') 
for daily closing oil prices of the brand West Texas Intermediate (WTI), a standard reference for American crude oil. The series, given in Figure \ref{WTI} below, contains 8497 values ranging from January 1986 up to September 2019. The  $p_\pi (d)$ are shown for the lags $d=1,...,10.$ The most apparent fact is that frequencies for different lags are very similar at each pattern. This will be discussed in the sequel.

On the left, frequencies are shown on linear scale. The largest frequencies appear for patterns 1234 and 4321, with $p_\pi (d)$ around $1/8.$ This happens for all daily price data. Here, $p_{1234}$ is clearly larger $1/8$ and  $p_{4321}$ is smaller, which can be explained by the increasing trend of the series. The largely increasing patterns 1243 and 2341 and their decreasing counterparts 4312, 3214 have probabilities near to $1/16=0.0625.$ All other patterns have smaller probabilities. 

On the right, pattern frequencies are shown on logarithmic scale, in order to distinguish better the rare patterns. Differences here actually represent quotients, and deviations from the mean have to be interpreted as relative errors.  The pattern with the smallest frequency, less then 1.5 percent, is 3142, followed by 2413 with almost 2 percent. Note that the average frequency is $1/24\approx 4.2\% .$ The patterns in the middle, with numbers 8 to 17, as well as 5,6 and 19,20, are below average.

The figure shows an obvious symmetry between patterns with number $k$ and $25-k.$ Actually, pattern number $25-k$
is obtained by reversing all order relations in pattern number $k.$ Rank 4 is interchanged with 1, rank 2 with 3, and $<$ with $>.$ The graph of pattern $25-k,$ drawn as in Figure \ref{ord3}, is the graph of the pattern with number $k$ turned upside down.  It seems plausible that these two patterns have the same frequency, but in our series it is not quite true. The permutations with higher number have somewhat smaller frequency than their counterparts, with exception of number 12 and 13: 3124 is more frequent than 2431. Can this break in symmetry be explained by a trend?

\begin{figure}[h] \centering
\includegraphics[width=.99\textwidth]{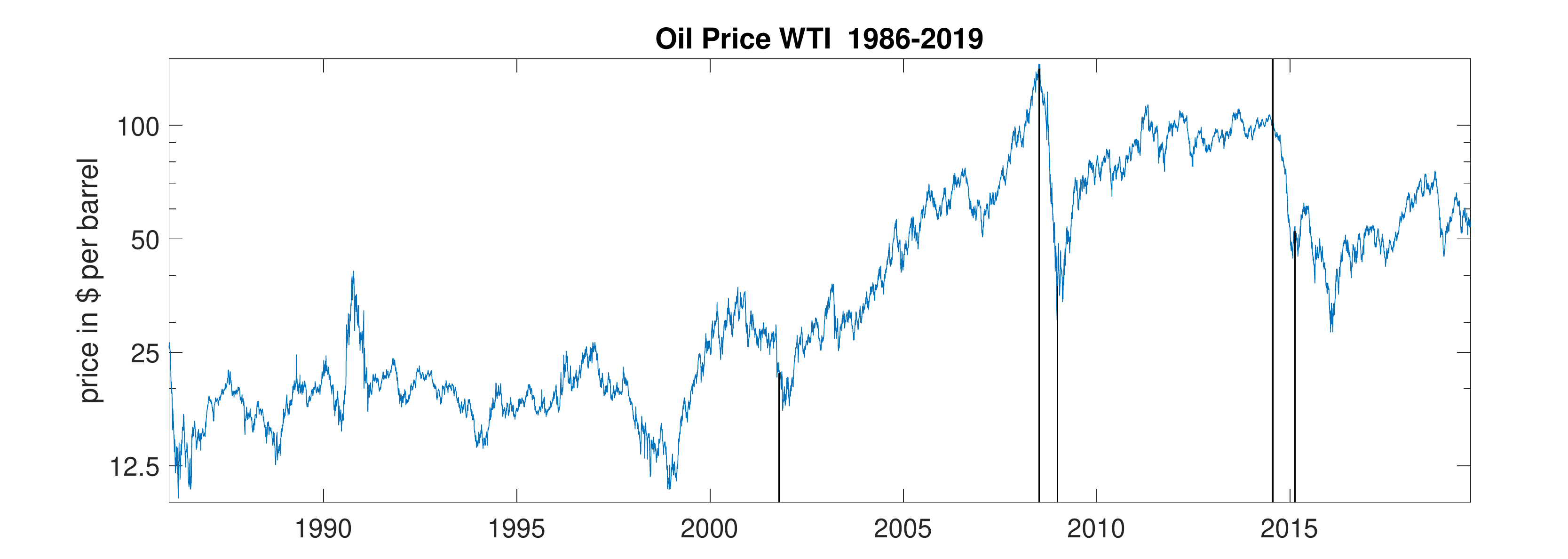}
\caption{The time series of WTI oil prices on logarithmic scale. The ad hoc segmentation contains two periods of increasing prices and two intervals without clear trend. Periods of decrease were too short to provide reliable pattern estimates.} \label{WTI}
\end{figure}  

The time series of $T=8497$ oil prices, drawn in Figure \ref{WTI} on logarithmic scale, shows a general upwards trend, as most price series do. It is interrupted by two rather sudden drops of prices: one in 2008 when the world economy was hit by the financial crisis, and one in 2014 when some producers, notably Saudi Arabia, increased their supply to compete with the upcoming fracking industry in the USA.  

So the series is not stationary and needs segmentation, which will be done by change point methods below. 
To study the effect of the trend on pattern frequencies, we construct an ad hoc segmentation into four pieces by eyesight.
The first subseries with 4000 values covers the time from 2 January 1986 to 16 October 2001 where no trend can be seen (the first price is 25.56\,\$, the last one only 22.01\,\$). The second series with 1680 values describes the time of strong price increase from 17 October 2001 up to 7 July 2008 with the maximum value. Another series with increasing trend and 1400 values covers the time from 26 December 2008 to 22 July 2014. The last series consists of the remaining 1147 values from 23 July 2014 to 3 September 2019, with almost no trend. The times of rapidly decreasing prices, with 120 business days in 2008 and 150 in 2014, were too short to provide a reliable statistics of pattern frequencies. 

\begin{figure}[h] \centering
\includegraphics[width=.7\textwidth]{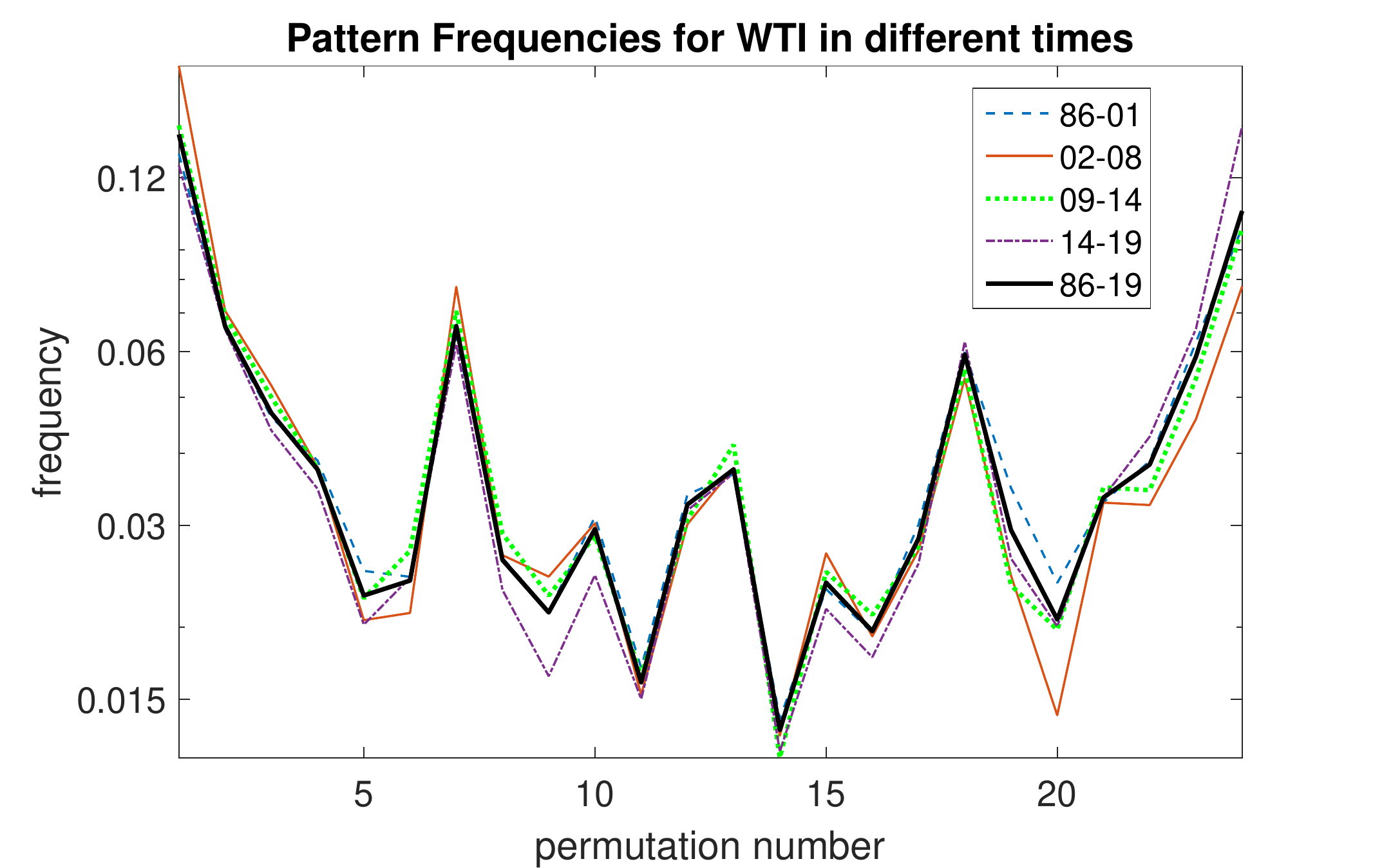}
\caption{The pattern frequencies of WTI in different time periods.} \label{WTIt}
\end{figure}  

Order four pattern frequencies for the four subseries are shown in Figure \ref{WTIt}.  They do not look as different as expected. It is true that for the strongly increasing series 2009-14, the increasing patterns on the left side of the figure have  larger frequencies and the decreasing patterns on the right side, with number 20 to 24, say, have smaller frequencies. And for the time 2014-19 without trend, pattern 4321 appears more often than 1234. But some of the frequencies, notably number 11 to 14 in the middle,  are almost constant through time. On the whole, frequencies do not differ much.

How can we decide about differences of such frequencies with statistical rigor?  This will be a theme of the next sections. First we introduce a standard model.

\section{Brownian motion as model for patterns in financial data} \label{brow}  
Since the work of Bachelier in 1900, Brownian motion (abbreviated BM) is a basic model for financial time series.  It is usually taken to describe the logarithm of prices. For day-to-day data it postulates that the difference $R_t=\log x_{t+1}-\log x_{t},$ the so-called log return of day $t,$ has a Gaussian distribution with mean zero and some variance $\sigma^2.$ Moreover, log returns of different days should be independent.

The independence assumption is plausible. When there would be a specific dependence between returns of today and tomorrow, some speculants would immediately bet against it to earn money, which would cause the dependence to disappear. The Gaussian distribution is not realistic, however, since big returns, both negative and positive, are more frequent in practice than the normal distribution allows \cite{ManF}. Moreover, the important phenomenon of varying volatility is not contained in the Brownian motion model.

For order patterns, Brownian motion is a good model. It makes no difference whether we model logarithms or original values, because $\log x$ is an increasing function. The value of $\sigma$ is also irrelevant, so we take $\sigma=1.$ Thus we let $x_t, t=1,...,T$ be a realization  of standard Brownian motion - the cumulative sum of a series of standard normal random numbers. It can be simulated by the Matlab command  x\,=\,cumsum(\,randn(1,T)\,).

\begin{figure}[h] \centering
\includegraphics[width=.7\textwidth]{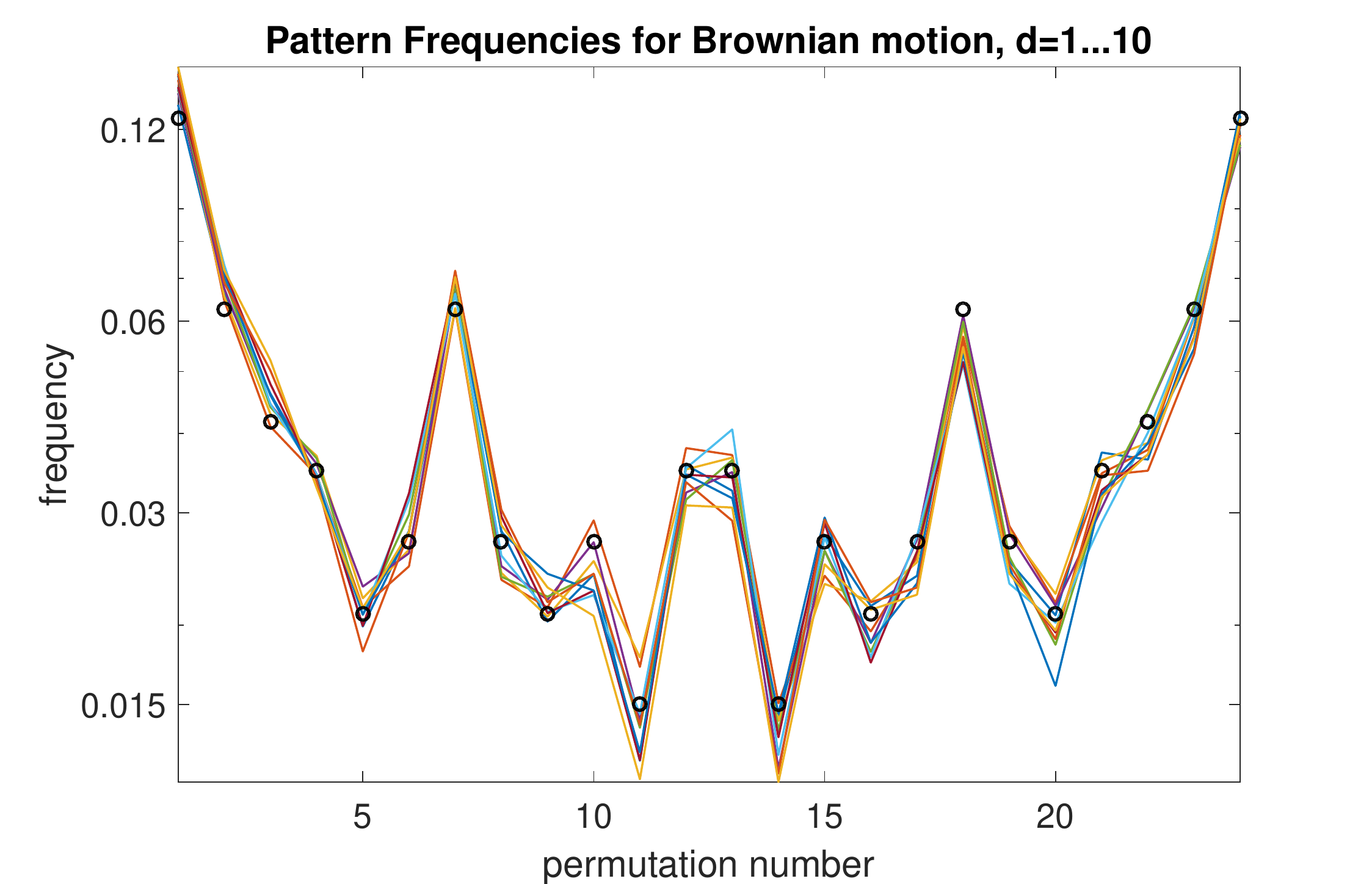}
\caption{A simulation of Brownian motion pattern frequencies for lags $d=1,...,10.$ We took $T=8500$ as for the WTI series above. Theoretical pattern probabilities are marked by circles. } \label{BM0}
\end{figure}  

Pattern frequencies for lags $d=1,...,10$ of a typical simulation of Brownian motion with length $T=8500$ are shown in Figure \ref{BM0}. Theoretically, patterns $k$ and $25-k$ appear with the same probability, but this need not be true in a simulation. Moreover, the process of Brownian motion in discrete time is known to be self-similar. More precisely,  $B_1,B_2,B_3,...$ and $(B_s,B_{s+d},B_{s+2d},...)/\sqrt{d}$ have the same distribution for any lag $d\ge 1$ and any initial time $s.$ As a consequence, the probabilities $p_\pi (d)$ do not depend on the lag $d.$ Of course, this need not be so in a simulation.  For large length $T=10^6$  we checked that the coincidence is convincing for virtually every simulation. Here we want to see what happens for $T=8500,$ the same length as our oil price data. Comparing Figures \ref{BM0} and \ref{WTI4}, we see that the coincidence of frequencies for different lags in the data is not worse than in the model, where it should be perfect  by definition. We try to clarify this point in the next section.

\begin{table}[h]
\centering
\begin{tabular}{|c|c|c|c|c|c|c|c|} \hline
Pattern number&1&2,7&3&4,12&5,9&6,8,10&11\\ \hline
Probability&$1/8$&$1/16$&$1/24$&0.035&$1/48$&0.027&0.015\\ \hline 
\end{tabular}
\caption{Exact probabilities of permutations of length 4 for Brownian motion. Patterns $k$ and $25-k$ have the same probability. Numbers with decimal point indicate irrationals involving $\arcsin 1/\sqrt{3}.$}\label{BM1}
\end{table}

For Brownian motion, probabilities of order 4 patterns can be determined analytically \cite{BS}: since the multivariate standard normal distribution is spherically symmetric, octant probabilities can be determined as surfaces of spherical triangles. The angles of these triangles are related to certain correlation coefficients. Table \ref{BM1} lists the results of \cite{BS}. It is curious that there are seven different probabilities of which only four are rational.

Comparing the distributions of pattern frequencies of WTI prices in Figure \ref{WTI4} and BM in Figure \ref{BM0}, we see little difference.  Is BM a good model in our case? We consider distributions of order 4 patterns as vectors in ${\mathbb R}^{24}$ and study their Euclidean distances. 

Let $q=(q_1,...,q_{24})$ denote an empirical distribution of WTI for one of the four time periods in Figure \ref{WTIt}. We determine the distance $\| q-\overline{q}\| $ to the distribution $\overline{q}$ of the WTI over the whole time, and the distance $\| q-b\|$ to the theoretical distribution $b$ of Brownian motion given in Table \ref{BM1}.  Moreover, $N=10^5$ simulations of BM with the sample size of the respective WTI series for $q$ are performed and their pattern distributions $b_k$ calculated. For  $k=1,..,N$ we determine $d_k=\| b_k-b\| .$ 

The distribution of all $d_k$ serves as test distribution for the null hypothesis that $q$ is obtained from the Brownian motion model. It resembles a chi-distribution for 24 degrees of freedom, with more heavy tail on the right, but not far from normal.   Our simulation provides the median of the $d_k,$ and an estimate of the $p$-value of the data: 
\[ p= \frac1N\, \# \{ k\, |\, 1\le k\le N\, ,\ d_k>\| q-b\|\, \} .\]

For different lags $d,$ the distances $\| q-b\|$ can differ by 20 or 50 \%, but they were always within a factor of 2. The resulting $p$-value, however, can differ by a factor 10 and more.  Thus for a distribution with 24 frequencies, the dependence on the lag can be a problem. This experience leads us to simpler parameters of the time series, introduced in section \ref{taube}.
 
Here we decided to take the mean of the frequencies for lags 1,2, and 3 as our empirical distribution $q.$ This method is justified in the next section.  The results are shown in Table \ref{BM2}. It turns out that for the last period 2015-19, BM is a perfect model with $p=88.9\% .$  The patterns of the third period 2009-14 could also well be generated from BM.  For the whole series as well as for the first two periods, there are significant differences, however.

\begin{table}[h]
\centering
\begin{tabular}{|c|c|c|c|c|c|} \hline
WTI time series&1986-2019&1986-2001&2001-08&2009-14&2015-19\\ \hline
Number of values&8497&4000&1680&1400&1150\\ \hline 
Distance $\| q-\overline{q}\|$ to whole WTI&0&.013&.047&.014&.014\\ \hline 
Distance $\|q-b\|$ to BM&.025&.024&.070&.034&.022\\ \hline 
Median of distances $\| b_k-b\|$&.011&.016&.024&.026&.029\\ \hline
\bf $p$-value in \% &\bf 0.04&\bf 3.8&$<$\bf 0.01&\bf 13.6&\bf 88.9\\ \hline 
\end{tabular}
\caption{Distances of pattern distribution $q$ of WTI prices in different periods to their mean $\overline{q}$ and to the exact distribution $b$ of BM. This was compared with the distance of $10^5$ simulated sample distributions $b_k$ of BM to $b.$ During the last two periods, Brownian motion fits well.}\label{BM2}
\end{table}

What is the reason for the differences? In the second period, it is the strongly increasing trend of the prices, which results in large frequencies $p_{1234}$ and small frequencies $p_{4321}.$  This is not surprising and could be mended by taking a biased BM with linear trend fitted to the increase of prices in that time. For the whole series, the patterns  would also fit better to a BM with linear trend. However, in that case the trend has to be chosen in such a way that today's prices would range above 300 instead of 60 \$. Thus there may be another structural difference between WTI and BM for the whole series and for the first rather stationary period, although the $p$-value of 3.8 \% in Table \ref{BM2} is not convincing.  We return to this question later.

\section{Order-self-similarity of financial data} \label{daily}
Before we go on, we make a few general remarks. Instead of studying the prices $x_t,$ it is common to consider the increments $Z_t=x_{t+1}-x_{t}$ or log returns $R_t=\log x_{t+1} -\log x_{t}$ of every day. Their distribution is determined, and usually has heavier tails than normal distribution. Here we study order patterns in the original data, however. Returns come in automatically when we compare values: $x_{t+1}>x_{t}$ means that the increment and log return of day $t$ are positive. Our order patterns can be considered as combinations of signs of returns for different days, and different time lags.

In contrast to time series which are measured with a fixed sampling rate, daily financial data come with an {\bf irregular time scale.} At weekends as well as on holidays, the stock market is closed. We pretend that Monday is the day following Friday, although we know that much business is going on throughout the weekend. Moreover, there are days where no trade takes place, and other days are very busy. It was suggested that the time scale should be modulated by the trading volume, but this is not done. As a consequence, studies of daily market data  by classical methods of time series analysis, as autocorrelation and frequency spectrum, do not reveal clear structure. 

On the positive side, it has often been observed that financial series have a nice symmetry property. They are statistically self-similar in the sense that a magnification of the function at one place looks very similar to the original function at another place \cite{ManF}. A random process in continuous time, $Y_t, t>0$ is said to be \emph{self-similar} if there is an exponent $H>0$ such that 
\begin{equation} 
Y_{rt}=r^HY_t \quad\mbox{in distribution, for each positive number } r. 
\label{self} \end{equation}
For details, see \cite{EM} or \cite{EKM}, section 8.9.
Brownian motion with $H=\frac12 $ is the standard example. Certain Levy Processes and fractional Brownian motion are also self-similar, and have been used as models of financial data. 
Usually, equation \eqref{self} is only considered for the one-dimensional distribution of the $Y_t,$ and there are methods
 to determine the Hurst exponent $H$ from data, see \cite{ManF,SK11}. The mathematical definition applies to the whole process, however. Thus for every $n,d\ge 1$
\begin{equation}
(X_{d},...   ,X_{nd})\ \mbox{ has the same distribution as }\ d^H\cdot (X_1,....,X_n)\ .
\label{sels}\end{equation}
This multidimensional setting is practically impossible to check with statistical methods.
Using order patterns, we now define a simpler concept.  Note that $(X_1,....,X_n)$ and $c\cdot (X_1,....,X_n)$ represent the same order pattern for each positive constant $c.$
 
So let us say that a process  $Y_1,Y_2,...$ is \emph{order self-similar} if for every $n,d\ge 1$
\begin{equation}
(X_{d},...   ,X_{nd})\ \mbox{ has the same pattern distribution as }\ (X_1,....,X_n)\ .
\label{selso}\end{equation}

{\bf Proposition 2. } \emph{Each self-similar process is order self-similar. If the process is also 
ergodic and has stationary increments, then $p_\pi=\frac1m\cdot\sum_{d=1}^m p_\pi(d)$ is a consistent and unbiased estimator of the pattern probability of $\pi ,$ for any fixed $m\ge 1.$} \vspace{2ex}

The proof follows from Proposition 1 and the remark above. Thus we can use order self-similarity to improve our estimators, as already done in the previous section.  Moreover, order self-similarity is a parameter-free concept - we do not need a Hurst exponent. And it seems rather easy to check this concept with sufficiently many data. \vspace{0.5ex}

{\bf Problem. } It is not known whether all order self-similar processes are self-similar. Note that self-similar processes are not stationary. The standard examples mentioned above have stationary increments, however. Are there stationary order self-similar processes?  \vspace{2ex}

Why should a daily financial series be self-similar?  There are small and big orders on the market. Some shareholders keep their items for a long time, others will sell them immediately.  Briefly, there are actions of all sizes which produce structure of all sizes in the time series.  This may be true, but in our opinion the irregular time sampling of daily financial series contributes a lot to their self-similarity. 

Our time series $x_t$ mixes time differences over one day, like Monday to Tuesday, with three-day differences from Friday to Monday. When we consider lag 3, like Monday to Thursday, then most differences cover five days, like Thursday to Tuesday.  With such an irregular scale, our only chance seems to postulate that order pattern probabilities do not depend on the lag, at least for small lags.  This postulate is supported by empirical data like Figure \ref{WTI4}. 

Henceforth, we shall assume that financial data fulfil order self-similarity in the sense of \eqref{selso}, for  $d\le 6.$ We use this assumption to improve estimates by applying Proposition 2  with $m=3$ or $m=6.$  It makes no sense to go to much larger $m.$ On one hand, the data will not fulfil \eqref{selso} for large $d,$ on the other hand the variance of estimates $p_\pi(d)$ increases with $d,$ as we shall see below.

\section{Turning rate and up-down balance} \label{taube}
Pattern frequencies are not so easy to interpret.  Certain combinations of pattern frequencies are more meaningful. Let us fix the lag $d=1$ for simplicity. The \emph{permutation entropy} of order $n$ is a sum over all patterns of length $n.$
\begin{equation}
H = - \sum_\pi p_\pi  \log  p_\pi  \ , 
\end{equation}
This was introduced as a measure of complexity of the data \cite{BP}. See \cite{Ami,ZZRP,AKK,AKU} for details, applications and related concepts. Here we shall focus on two simpler quantities. For patterns of length 2, we consider the
\begin{equation}
\mbox{\it up-down balance }\qquad \beta = p_{12}-p_{21}= 2p_{12}-1 \  .\label{beta}
\end{equation}
This parameter measures asymmetry, non-Gaussianity or irreversibility of the process \cite{BS,Ba15,Ba19}. If there is an increasing or decreasing trend in the series, then $\beta$ will be positive or negative, respectively. In the extreme case of an increasing or decreasing time series $\beta$ assumes its maximum 1 and minimum $-1,$ respectively.
However, $\beta$ can also be positive in a stationary process. For instance, a stock price could increase always from Monday to Friday, and then fall to the level of last Monday over the weekend. In that case, $\beta =\frac45-\frac15=\frac35 .$  Many time series in nature, for example the daily number of sunspots \cite{BS}, show similar behaviour.
A random walk, that is, a process with independent stationary increments, will have no trend if the increment $Z=X_{t+1}-X_t$ has zero mean. The median need not be zero, and in that case $\beta =2P(Z>0)-1$ will be positive or negative. There are also asymmetric Levy processes where the mean of $Z$ does not exist. An autoregressive process with exponential noise is stationary and has nonzero $\beta$ for small $d.$ An example is discussed at the end of this paper. For financial data, $\beta$ often differs from zero, but not too much, as Table \ref{albe} indicates for our oil data.

Our other important parameter is based on patterns of length 3. It is the
\begin{equation}
\mbox{\it turning rate }\qquad \alpha = p_{132}+p_{231}+ p_{213} +p_{312}=1- p_{123}-p_{321} \ ,  
\end{equation}
the relative amount of local maxima and minima (turning points) in the series.  This is an intensity parameter which varies between 0 and 1. The extreme case zero is approached by a smooth time series which has very few turning points. The other extreme is an alternating series, like $x_t=(-1)^t,$ for which each point is a turning point. For a series of independent random numbers (white noise) we have $\alpha=\frac23 ,$ and for Brownian motion $\alpha=\frac12 .$

In previous work \cite{BS,Ba15,Ba19}, the author considered the persistence $\tau =p_{123}+p_{321}-\frac13 =\frac23 -\alpha$ instead of $\alpha$ as a function of $d.$ This function shares many properties with autocorrelation. It is zero for white noise, for instance. However, turning rate was essentially introduced in \cite{bien1} and is a very intuitive concept while the word `persistence' is used with various meanings even in mathematics. So $\tau$ is not used here.

$H, \beta ,$ and $\alpha$ can be considered for arbitrary lag $d\ge 1.$ So they become functions of $d$ which can be used as correlation functions. This is very useful in the case of big data series from weather or medicine where certain periodicities appear on various scales \cite{Ba19}. The result shown in Figure \ref{sl1} was obtained with the turning rate for $d=4.$ For financial data, which are approximately order self-similar, we have essentially only one value $\alpha$ for small $d.$  But Proposition 2 provides different ways of calculation, in particular
\[ \alpha^{1}=\alpha (1)\ ,\qquad  \alpha^3=\frac13 \sum_{d=1}^3 \alpha  (d) \ ,\qquad\mbox{and }\quad
\alpha^6=\frac16 \sum_{d=1}^6 \alpha  (d) \ . \] 
The same holds for $\beta .$ In Table \ref{albe} we compare $\alpha^1,\beta^1.$ with $\alpha^3,\beta^3.$ We see that the turning rate of our oil data is $\frac12$ in all periods, with very small fluctuation. The differences of $\beta$ from zero are a bit larger. The $z$-values are obtained from Proposition 3 below.

\begin{table}[h]
\centering
\begin{tabular}{|c|c|c|c|c|c|} \hline
WTI time series&1986-2019&1986-2001&2001-08&2009-14&2015-19\\ \hline
Number of values&8497&4000&1680&1400&1150\\ \hline 
turning rate $\alpha^1$&0.510&0.506&0.528&0.509&0.507\\ \hline 
turning rate $\alpha^3$&0.502&0.512&0.488&0.500&0.499\\ \hline 
$z$-value of $\alpha^1$&1.84&0.76&2.29&0.67&0.47\\ \hline 
up-down balance $\beta^1$&0.032&0.025&0.072&0.044&0.050\\ \hline 
up-down balance $\beta^3$&0.044&0.038&0.112&0.063&0.035\\ \hline 
$z$-value of $\beta^1$&2.95&1.58&2.95&1.65&1.69\\ \hline 
\end{tabular}
\caption{Comparison of two estimators of turning rate and up-down balance for different segments of the oil price series.
$z$-values were calculated from Proposition 3 below.}\label{albe}
\end{table}

\section{The variance of $\alpha$ and $\beta$} \label{vari}
Defined in terms of the estimators $p_\pi,$ the quantities $H, \alpha,$ and $\beta$ are in fact themselves estimators of respective parameters of the underlying process. For instance, $\beta$ estimates $b=P(X_1<X_2)-P(X_1>X_2).$  For an ergodic process with stationary increments, Proposition 1 implies that $\alpha$ and $\beta$ are consistent and unbiased estimators. The entropy $H,$ given by a continuous but nonlinear function of the $p_\pi,$ is consistent but not unbiased.

Now we determine the quadratic error of these estimators, at least for the case of Brownian motion. In other words, we calculate the variance of the functions $\alpha$ and $\beta ,$ taken over all possible realizations of BM. We start with a basic fact. \vspace{2ex}

{\bf Proposition 3. } \emph{For samples of length $T$ from Brownian motion and lag $d=1,$ the number of turning points and the number of up-steps both follow the statistics of tossing a fair coin. The distribution of the number of turning points is binomial with $n=T-2$ and $p=\frac12 .$ The number of $t$ between 1 and $T-1$ with $x_t<x_{t+1}$ has binomial distribution with $n=T-1$ and  $p=\frac12 .$ \quad As a consequence, we have}
\[ E(\alpha)=\frac12\ ,\quad Var(\alpha)=\frac{1}{4(T-2)}\ , \quad E(\beta)=0\ ,\quad Var(\beta)=\frac{1}{T-1}\ . \] \vspace{2ex}

\emph{Proof. } The increments $Z_t=X_{t+1}-X_t$ of BM are independent and have positive sign with probability $\frac12 .$ Thus counting the number $n_{12}$ of up-steps $t$ with $Z_t>0$ is really the same as tossing $T-1$ fair coins and counting heads. Then $p_{12}=n_{12}/(T-1)$ has mean $\frac12$ and variance $\frac{1}{4(T-1)},$ and the assertion for $\beta=2 p_{12}-1$ follows.\\
A point $t$ between 2 and $T-1$ is a turning point (local maximum or minimum) of the time series $x_1,...,x_T$ if the increments $z_{t-1}=x_t-x_{t-1}$ and $z_t=x_{t+1}-x_t$ have the same sign.  The events to be counted are now
$A_t=\{ Z_{t-1}Z_t>0\}$ for $t=2,...,T-1 .$ Since the signs of the $Z_t$ are independent and are $+$ with probability $\frac12,$ we have $P(A_t)=\frac12$ for each $t.$ When we show that the $A_t$ are independent, we do again have a series of fair coin tosses. From the definition of $A_t,$ it is clear that $A_t$ is independent on all $A_s$ with $s\le t-2.$ It remains to show that $A_{t-1}$ and $A_t$ are independent: 
\[ P(A_{t-1}\cap A_t)=P( Z_{t-2},Z_{t-1}\mbox{ and }Z_t\ \mbox{ have the same sign})=
\textstyle\frac14=P(A_{t-1})P(A_t)\, .
\]  The rest of the proof is similar as for $\beta . \hfill \Box$ \vspace{2ex}

We apply the result to $\alpha^1$ and $\beta^1$ in Table \ref{albe}, to test the fit of Brownian motion to our oil price data. The $z$-value of $\alpha^1$ is obtained as 
\[ \textstyle
z=(\alpha -\alpha_{BM})/\sigma(\alpha_{BM})= (\alpha -\frac12)\cdot 2\sqrt{T-2} ,
\]
for $\beta^1$ we calculate $z=\beta \sqrt{T-1} .$ We compare with the standard normal distribution. For turning rates,  only the period 2001-08 showed significant differences. The $z$-values of $\beta$ are larger, but the significance for 2001-08 and for the whole time series can be attributed to trend. On the whole, BM is a good model for order patterns of the oil prices. 

Proposition 3 is paradigmatic because it shows that for all estimators $p_\pi (d)$ we can expect standard deviation and confidence bound of the form $C/\sqrt{T}$ where $C$ is a constant. However, beware of correlations!  The proposition is not true for lags $d$ greater than 1.  The reason is that $Z_t(d)=X_{t+d}-X_t$ and $Z_{t+1}(d)=X_{t+1+d}-X_{t+1}$ are not independent since contain the common part $X_{t+d}-X_{t+1},$ and this term is likely to become the main term when $d$ is large. So there is positive correlation among the sets $A_t,$ which increases the variance. For BM this can be studied rigorously but it is easier to perform a simulation. 

\begin{figure}[h]  
\centering
\includegraphics[width=.9\textwidth]{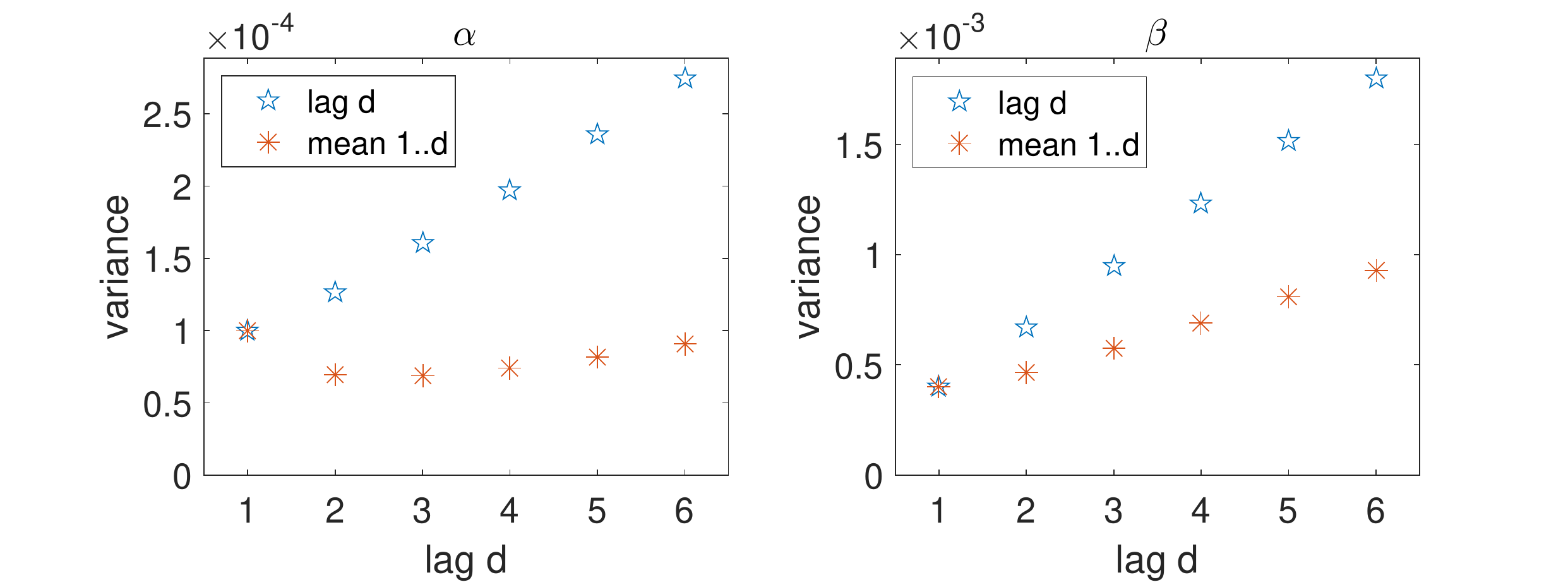}
\caption{Dependence of the variance of $\alpha$ and $\beta$ on the lag $d$ for BM, obtained by simulating $10^5$ trajectories of BM with length $T=2500.$. The variance of $\alpha$ increases slowly, and an average over lags 1 to 3 will diminuish the variance of the estimator. For $\beta$ the variance increases even in the average.} \label{abevar}
\end{figure}  

The result, shown in Figure \ref{abevar}, is surprising. The variance of both $\alpha$ and $\beta$ increases almost linearly with $d.$ For $\alpha$ the slope of the line is about $\frac13 ,$ and the average $\alpha^3$ has much smaller variance than $\alpha(1).$ For $\beta ,$ however, the slope is about $\frac45 .$ The increase is so strong that averaging over different lags does not improve the accuracy of the estimator.\vspace{1ex}

We have no space to extend this discussion, but for historical reasons we should mention the groundbreaking result of \cite{bien1}. He considered turning points for white noise as model for series of astronomical measurements \cite{bien2}. His 1874 paper was one page without proof. On the last page of his second paper, it is noted that J. Bertrand presented a proof on the next session of the French society for mathematics and astronomy. We slightly modify the result and add the statistics of up-steps.\vspace{2ex}

{\bf Bienaym\'e's theorem. } \emph{Consider a sequence of $T$ independent and identically distributed random variables. The number $V$ of turning points is asymptotically normal with} 
\[  M(V)=\frac23 (T-2)\qquad\mbox{ and }\quad Var(V)=\frac{8}{45}(T-2) +\frac{1}{30}\ . \] 
\emph{The number $U$ of up-steps is asymptotically normal with} 
\[  M(U)=\frac12 (T-1)\qquad\mbox{ and }\quad Var(U)=\frac{1}{12}(T-1) +\frac{1}{6}\ . \] 

\emph{Proof. } In an i.i.d. sequence of length $m,$ all permutations of length $m$ are represented with the same probability. This will be the main argument. 

We start with up-steps. Note that $U=\sum_{t=1}^{T-1} U_t$ where $U_t={\bf 1}_{\{ X_t<X_{t+1}\} }$ denotes the indicator function of up-step $t.$ Since $M(U_t)=p_{12}=\frac12$ for each $t,$ we get $M(U)=\frac12(T-1).$ The variance $Var(U)=M(U^2)-M(U)^2$ is expressed as the sum
\[ \sum_{t=1}^{T-1}\sum_{s=1}^{T-1}(M(U_s U_t)-{\textstyle\frac14}) = \sum_{t=1}^{T-1} (M(U_t^2)-{\textstyle\frac14}) +2\sum_{t=1}^{T-2} (M(U_t U_{t+1})-{\textstyle\frac14})\ .\]
We used that $M(U_sU_t)=\frac14$ for $|s-t|>1$ because of independence. Now $M(U_t^2)=\frac12$ and 
$M(U_t U_{t+1})=p_{123}=\frac16$ implies
\[ Var(U)=\frac{T-1}{4} -2 \cdot\frac{T-2}{12}= \frac{T+1}{12}\ .\]
For turning points, let $V_t={\bf 1}_{\{ t \mbox{ is turning point}\} }$ for $t=2,...,T-1.$ Then $M(V_t)=p_{132}+ p_{213}+p_{231}+p_{312}=\frac46$ implies that $V=\sum_{t=2}^{T-1}V_t$ fulfils $M(V)=\frac23(T-2).$\vspace{0.5ex}
 
To expand  $Var(V)=M(V^2)-M(V)^2$ as a sum, we now have to subtract $\frac49$ instead of $\frac14 ,$ and we can neglect terms with $|s-t|>2$ for which $V_s$ and $V_t$ are independent. We obtain $M(V_tV_{t+1})=10/24$ by looking at Table \ref{lexi} and realising that 10 of the 24 permutations have two turning points in the middle. In terms of the covariance we get  $Cov(V_t,V_{t+1})=M(V_tV_{t+1})-\frac49=-1/36 .$

It takes a little longer to see that exactly 54 of 120 permutations of length 5 have turning points at their second and second-to-last position. This implies that $Cov(V_t,V_{t+2})=M(V_tV_{t+2})-\frac49=1/180 .$ Then
\[ Var(V)=(T-1) Var(V_t) +2(T-2)Cov(V_t,V_{t+1})+2(T-3) Cov(V_t,V_{t+2})\ \]
gives the result after a brief calculation.  Asymptotic normality for $T\to\infty$ should be rigorously formulated in terms of standardized random variables. In Bienaym\'e's time it was taken for granted. Today it follows from central limit theorems for $k$-dependent or strongly mixing sequences of random variables. 
\hfill $\Box$ \vspace{2ex}

Note that the variances here are even smaller than for the binomial distribution in Proposition 3, because of negative correlations.  This often happens for the variance of pattern frequencies $p_\pi$ since many patterns, for instance 132, are not compatible with their shifted pattern.  Moreover, Bienaym\'e's theorem holds for any lag $d,$ in contrast to Proposition 3.  But now let us turn to change points.

\section{Change points with respect to mean and order structure} \label{chang}
Certainly there is no unique way to determine change points in a financial time series. From the mathematical point of view, the simplest way is to look for changes in mean.  A classical method takes the time point $k$ where the difference of the means $m_k= \frac1k \sum_{t=1}^k x_t$ and $\tilde{m}_k=\frac{1}{T-k} \sum_{t=k+1}^T x_t$ assumes a maximum or minimum. Thus we maximize the function 
\begin{equation}
f(k)= c_k \left| m_k- \tilde{m}_k  \right| \quad\mbox{ with }\ c_k= 2\sqrt{k(T-k)}/T \ .
\label{meac}\end{equation}
The factor $c_k$ normalizes the standard deviation of the difference of means, as in a two-sample $t$-test. 
(We did not estimate the standard deviation  since that would require information on autocorrelation.) We put the constant 2 into $c_k$ so that the factor is 1 for all $k$ in the middle of the interval, and $f(k)$ can be interpreted as a difference.  

\begin{figure}[h]  
\centering
\includegraphics[width=.32\textwidth]{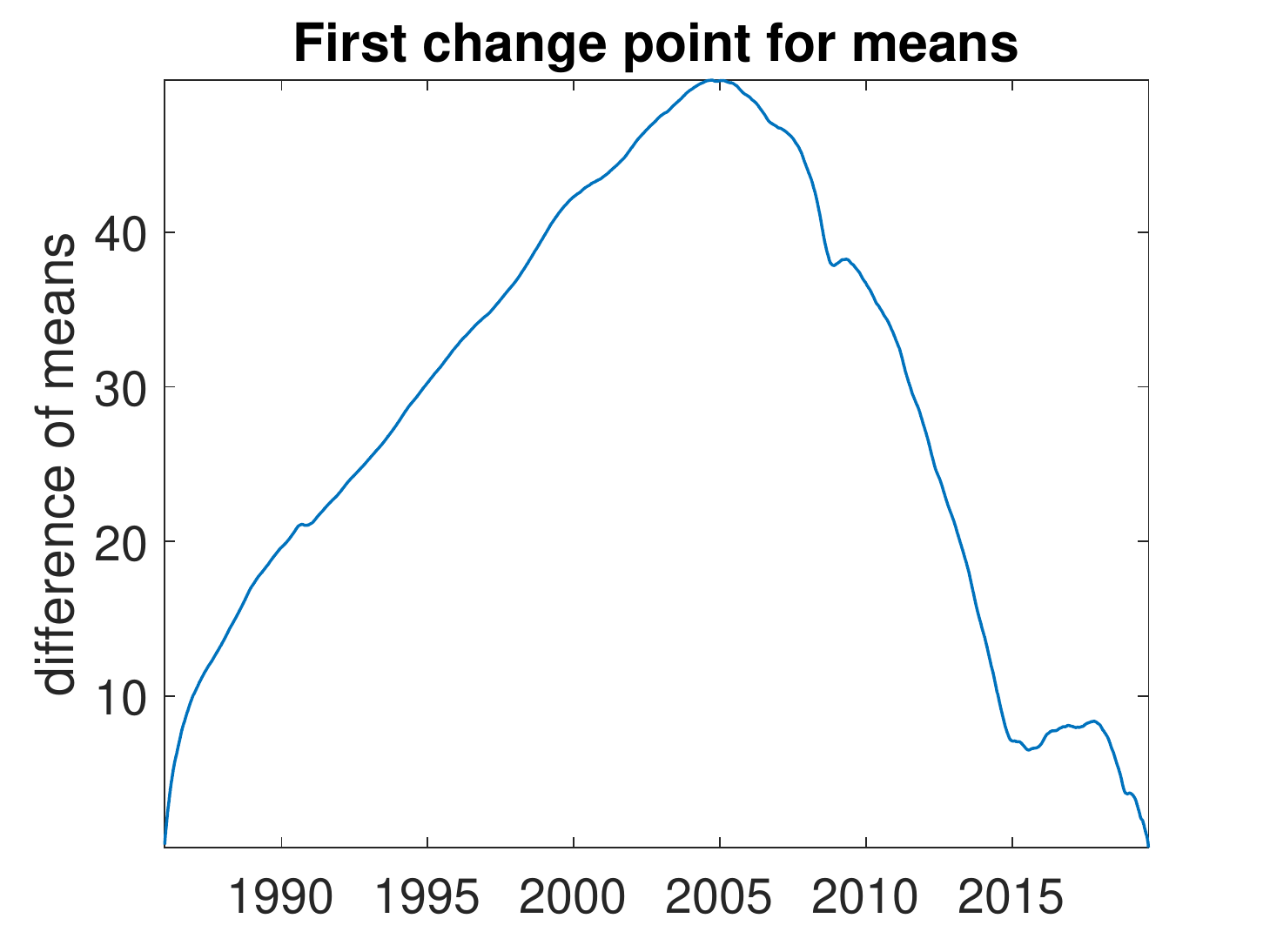} 
\includegraphics[width=.32\textwidth]{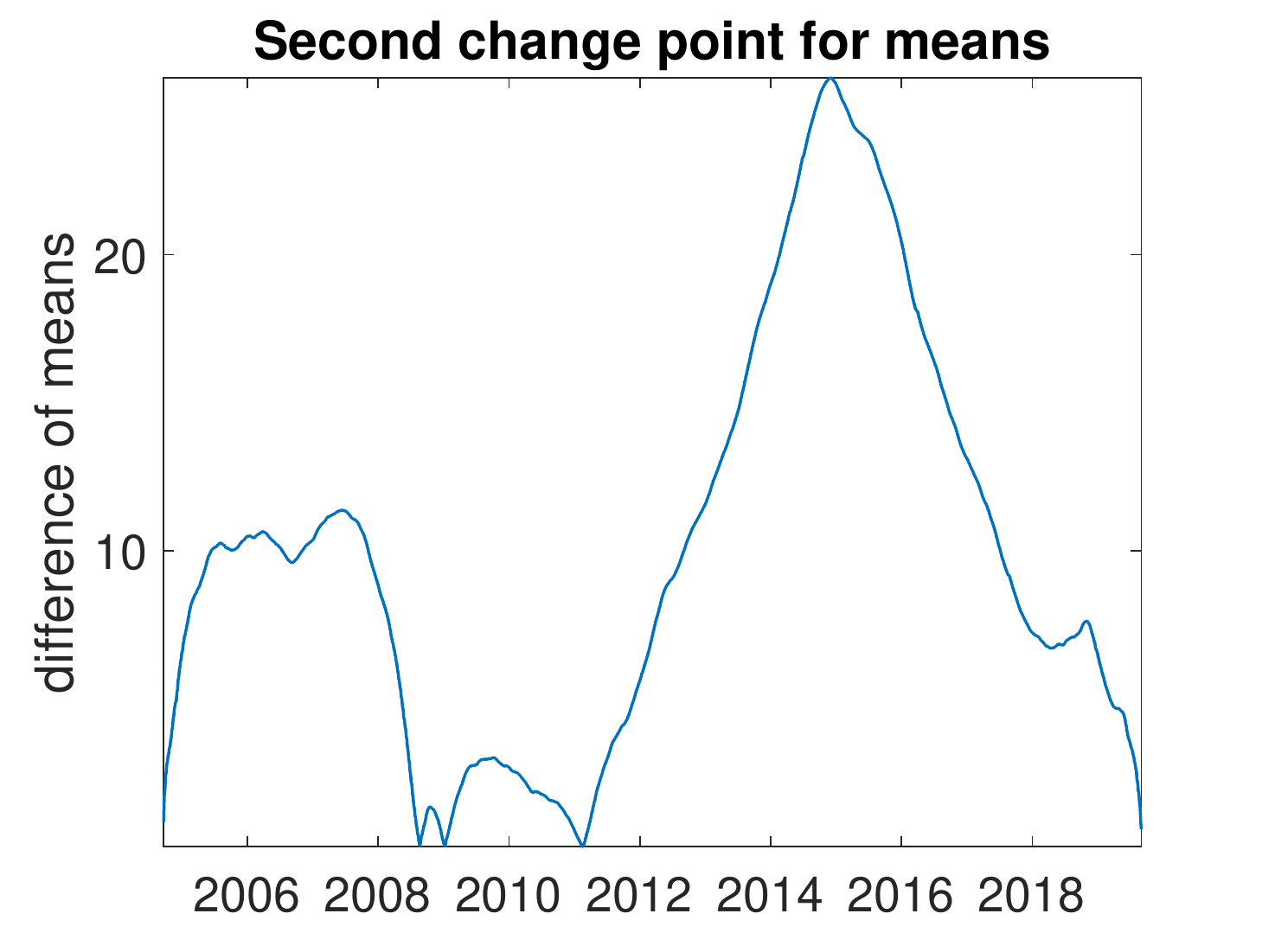}
\includegraphics[width=.32\textwidth]{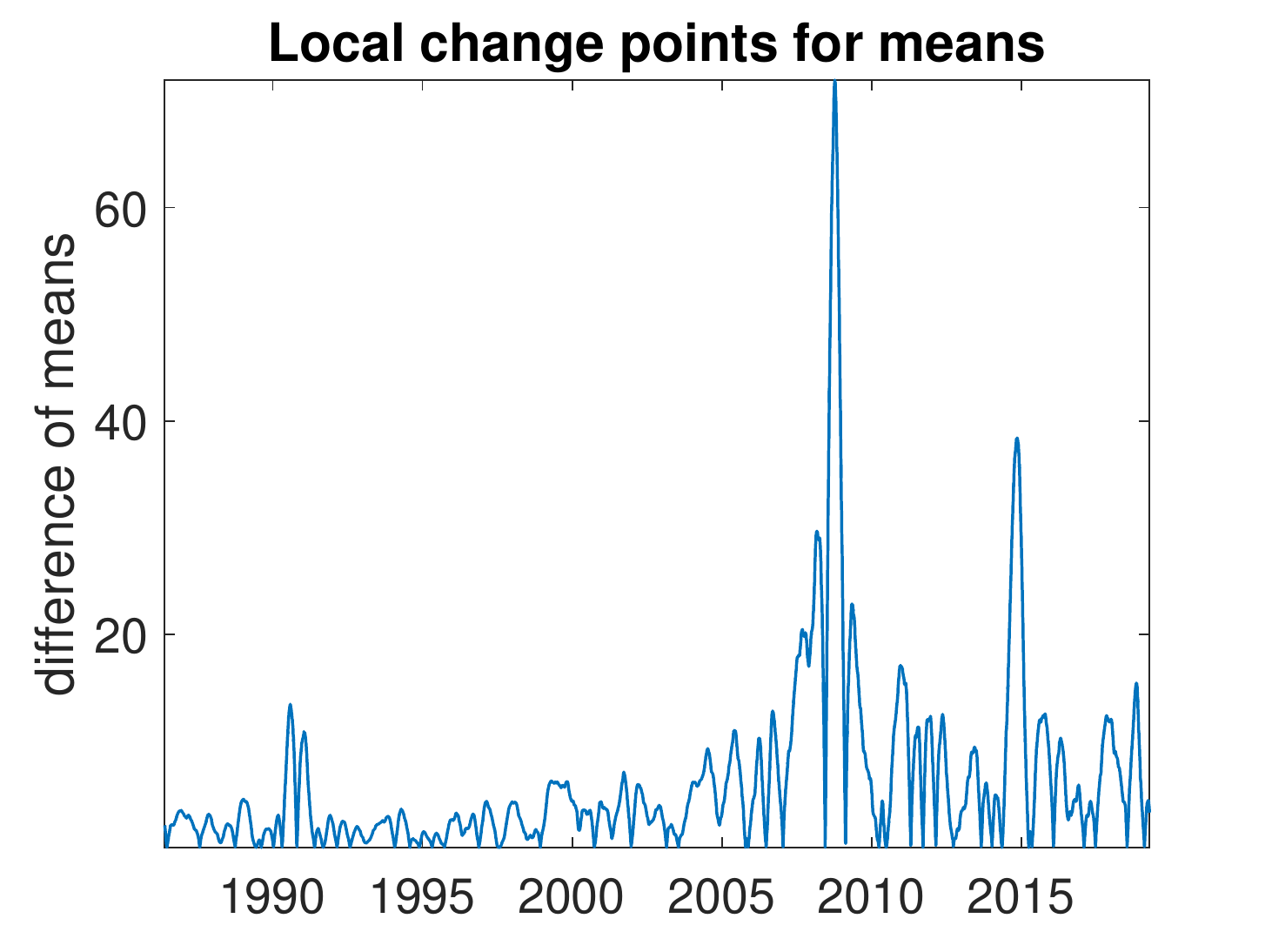}
\caption{First and second change point, and local change points in the mean for the oil prices.}\label{means}
\end{figure}

As seen in Figure \ref{means}, the function $f$ for the WTI series is rather smooth and has a unique maximum point $k_1$ in the strictly increasing part in the middle of the series in Figure \ref{WTI},  in late 2004.  All $x_t$ with $t<k_1$ are smaller, and almost all $x_t$ with $t>k_1$ are greater than the price at $k_1.$ The average difference is 50 \$. This is a clear solution to the optimization problem.  However, it is not a point where change happens.

When the method is applied to the time period between $k_1$ and $T,$ we again obtain a clear maximum at the end of 2014. It separates the time of very high prices before 2015 from medium prices we have had since then. Situated on the middle of a rapidly decreasing branch, this is again not a point where change happens.

\begin{figure}[h]  
\centering
\includegraphics[width=.9\textwidth]{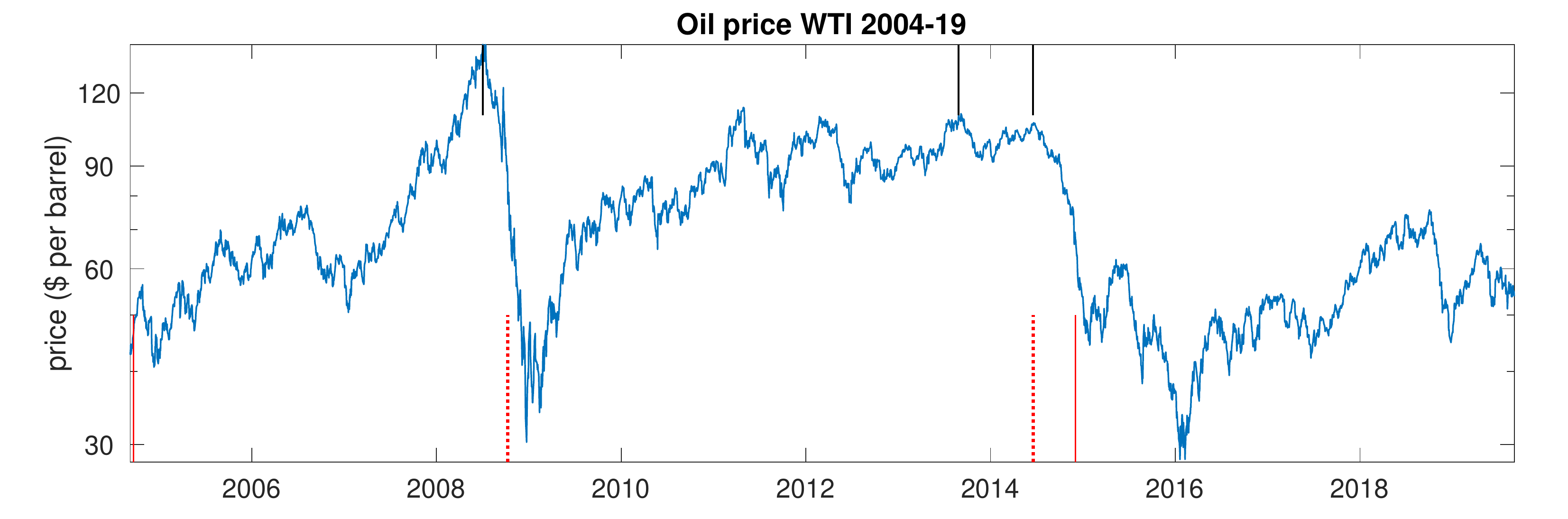}
\caption{The oil price series from 2004 to 2019. Calculated change points in mean are indicated at the bottom, local change points by dotted line. Change points in order are indicated at the top.} \label{meanloc}
\end{figure}  

We can also detect change in the mean locally, by comparing the last 100 points before $x_k$ with the mean of the next 100 points. As shown in Figure \ref{meanloc}, this results in two clear maximum points. One is in the middle of the 2008 period of rapid decrease. The other one at the beginning of decrease in 2014 seems a real change point. A lot of smaller maxima indicate smaller periods of increase and decrease.

Now we turn to order structure. Let $q^k=(q_1,...,q_{24})$ denote the vector of order 4 pattern frequencies for the time interval $[ 1,k],$ and $\tilde{q}^k$ the respective vector for the time period $[k+1, T].$ Let $\| v\|$ denote the Euclidean norm of a vector $v.$ Similar as in \cite{SKC}, we consider the function
\begin{equation}
g(k)= c_k\ \| q^k-\tilde{q}^k\| 
\label{ordc}\end{equation}
where the constant $c_k$ is the same as in \eqref{meac}. It serves to diminuish the bad estimates of frequencies for $k$ near the marginal values 1 and $T,$ and can be justified by the $1/\sqrt{T}$ magnitude of the standard deviation of the pattern frequencies discussed above.  Maxima of $g$ are taken as change points for order.  

\begin{figure}[h!]  
\centering
\includegraphics[width=.8\textwidth]{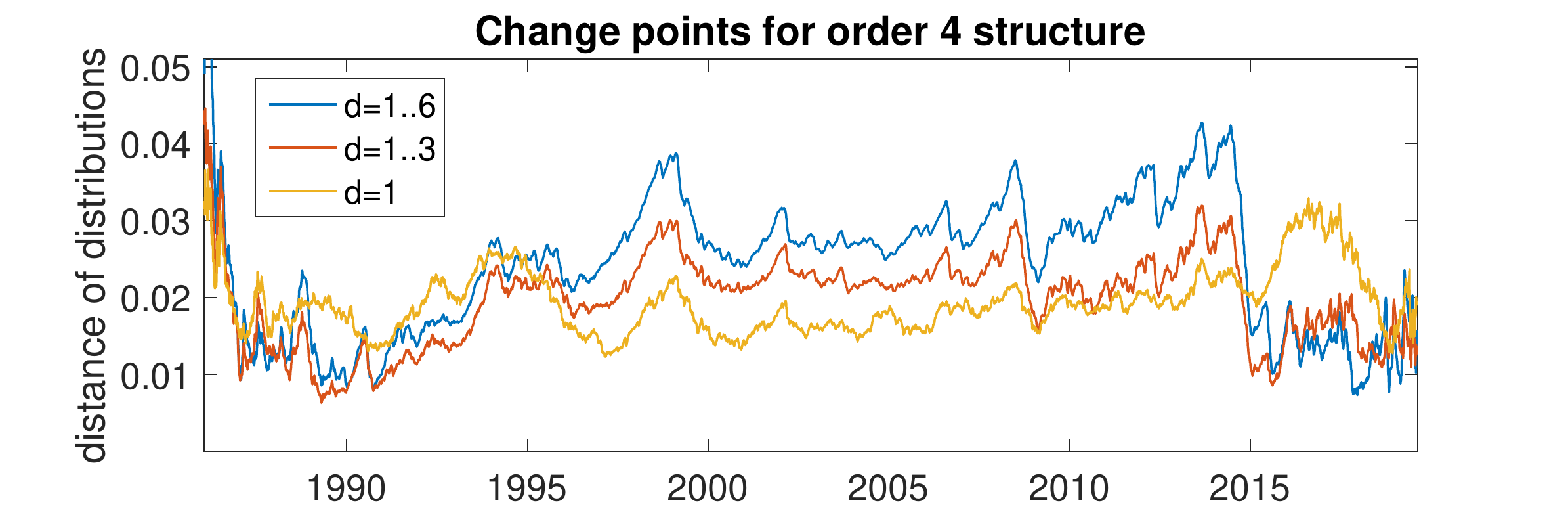} \\
\includegraphics[width=.8\textwidth]{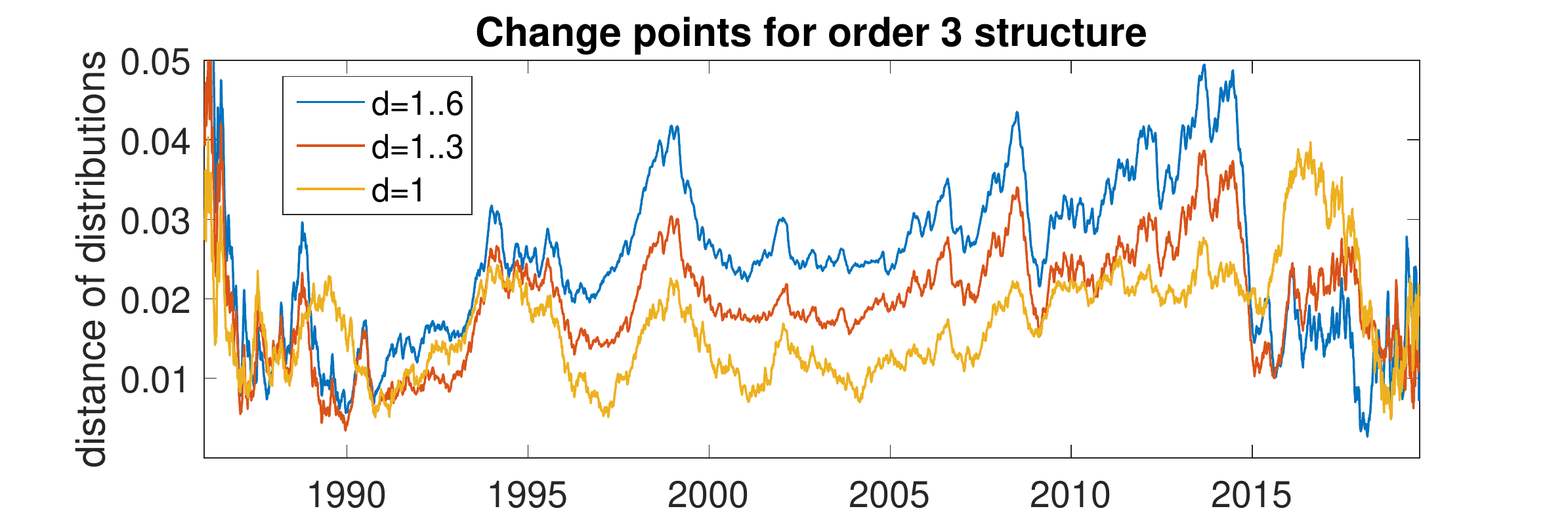}
\caption{Distance of pattern distributions before and after time $k.$ Means of lags give better results, while patterns of length 3 or 4 lead to the same change points.} \label{ordch}
\end{figure}  

The function $g(k)$ is shown in three versions in the upper panel of Figure \ref{ordch}. For lag $d=1,$ no clear change points are found. The maximum is near the right margin and turns out to be a random effect too large to be eliminated by $c_k.$ When we take $q$ from a mean of lags, 1 to 3, or 1 to 6, we see clear maxima at the end of 1998, in 2008 and 2013/14. The maximum at the left margin is again a random effect due to high variance of $g$ for small $k.$ On the right, the maximum found for $d=1$ has disappeared. 

When we compare with the time series in Figures \ref{WTI}, we see that the peaks in July 2008 and July 2014 were included in our ad hoc segmentation. The peak in December 1998 is also a convincing change point. Probably it is a better choice than our first segmentation point in October 2001. And the peak in August 2013 is also reasonable. It could be taken as an alternative to July 2014, as can be seen in Figure \ref{meanloc}. Moreover, the two means of lags give almost identical results, with few days difference in their maximum points. Distances of means over 1 to 6 are on a larger level than those from means over 1 to 3, because of larger variance discussed at Figure \ref{abevar}.

In the lower panel of Figure \ref{ordch}, the same results are shown for patterns of length 3. Again, $d=1$ is not a good option.  It was surprising that for means over lags 1 to 3 and 1 to 6 we get the same structure of maxima as for length 4. This indicates that with respect to change points the six patterns of length 3 contain all relevant information. 

\begin{figure}[h]  
\centering
\includegraphics[width=.7\textwidth]{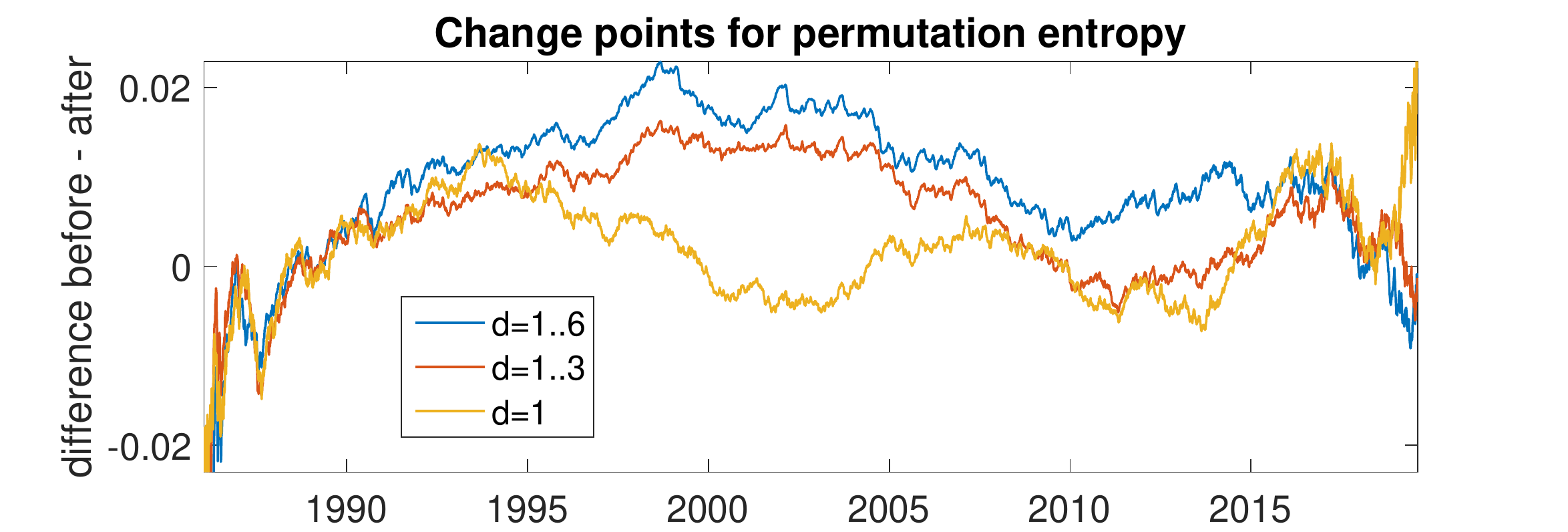}\\
\includegraphics[width=.7\textwidth]{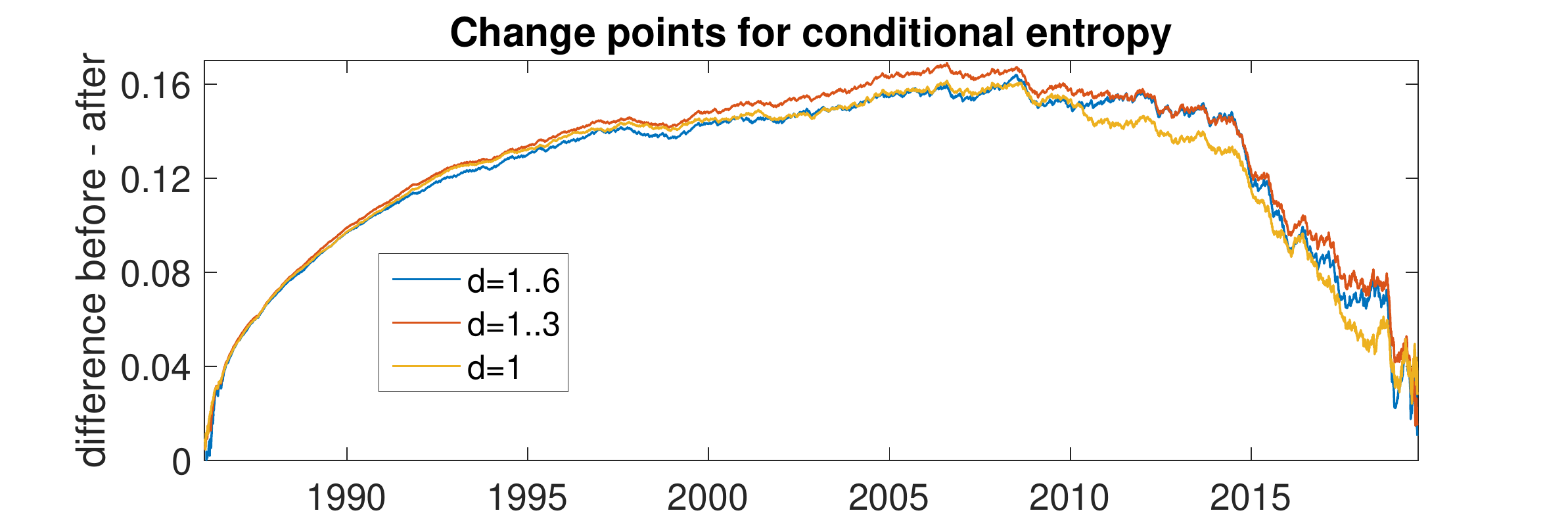}\\
\includegraphics[width=.7\textwidth]{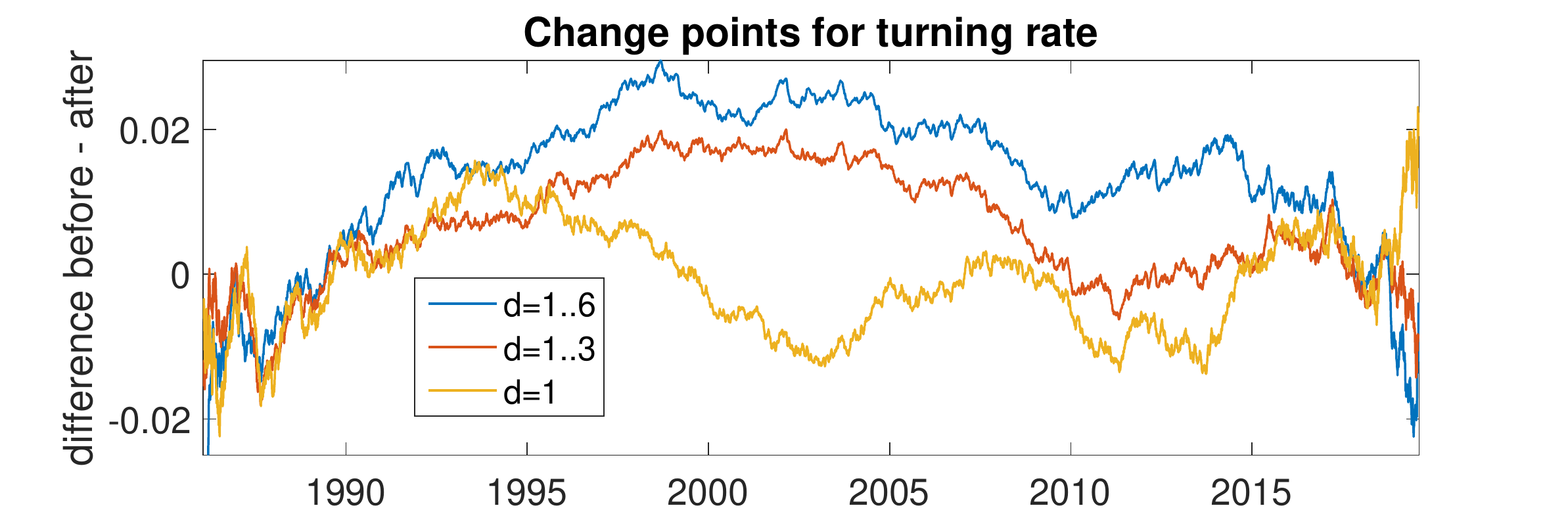}
\caption{Difference before and after $k$ of order 3 permutation entropies, conditional entropies recommended by \cite{UK18}, and turning rates. These methods seem not appropriate for segmentation of financial data.} \label{ordet}
\end{figure}  

Can we simplify even further?  There are three objectives: the function $g$ should become less erratic, more like the function $f$ for the mean. The criterion should be transparent, interpretable, as it is for the mean.  And the estimates of structure before and after should be fairly accurate even for small $k$ or $T-k.$ This was not the case for \eqref{ordc} where 1000 values would be minimal for a reasonable estimate of all 24 pattern frequencies. Even with $c_k$ there are marginal effects in Figure \ref{ordch}.

In the upper panel of Figure \ref{ordet}.
we tested the function $h(k)=c_K(H_k-\tilde{H_k})$ where $H,\tilde{H}$ denote permutation entropy of order 3 before and after $k,$ respectively.  The maximum in July 1998 is a realistic change point, but it is not convincing.  In the middle panel we used conditional permutation entropies recommended by \cite{UK18}. They work well for time series from certain dynamical systems with a Markov structure.  We have the same function $h(k)$ with  $H=\sum_{i=1}^4 -p^i\log p^i + p\log p +(1-p)\log 1-p$ where $p=p_{12}, p^1=p_{123}, p^2=p_{132}+p_{231}, p^3=p_{213}+p_{312}$ and $p^4=p_{321}.$ The function is rather smooth but there is no clear maximum. Our data do not include a Markov structure.
In the lower panel we used turning rates, $h(k)=c_k(\alpha_k-\tilde{\alpha_k}).$ As noted in Section \ref{slee} they worked so well for classifying sleep brain data. Here they give almost the same result as permutation entropy.  All three functions seem not appropriate for segmenting financial data.

\section{Change points with respect to up-down balance} \label{changup}
\begin{figure}[h]  
\centering
\includegraphics[width=.8\textwidth]{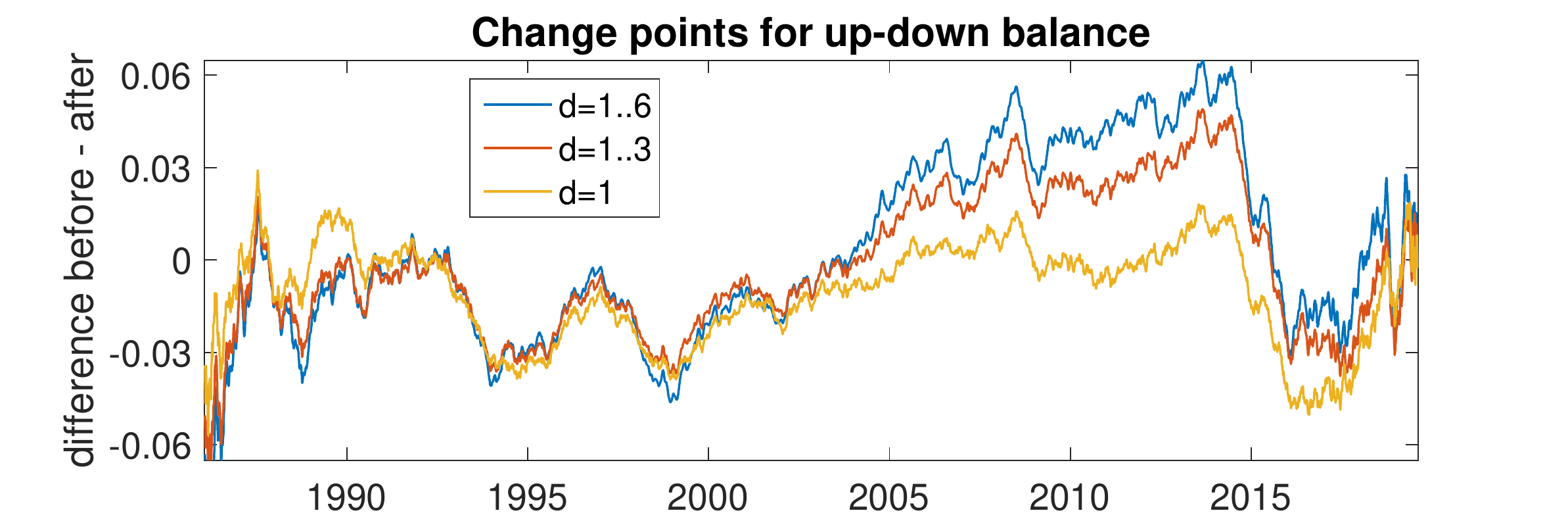}
\caption{Difference of $\beta$-values before and after $k.$ Up-down balance seems the most simple and most relevant parameter for change points in financial time series.} \label{ordb}
\end{figure}

Our last trial is the function $\beta .$ It will give the best results. We look for maxima and minima of $h(k)=c_k(\beta_k-\tilde{\beta_k})$ where $\beta_k$ and $\tilde{\beta_k}$ are the values of $\beta$ on $[1,k]$ and $[k+1,T],$ respectively.
Note that by Proposition 3, $c_k$ is exactly the normalizing constant for $\beta$ and $\alpha ,$ at least for lag 1.
Figure \ref{ordb} shows that lag $1$ alone is sufficient only from 1992 to 2002. The other two estimators of $\beta $ give very similar functions. Their main maximum and minimum points are very near to the maxima for the order structure in Figure \ref{ordch}.  The first change point $k_1$ has to be taken in August 2013. Our ad hoc point in July 2014 would be a possible alternative choice.

\begin{figure}[h]  
\centering
\includegraphics[width=.63\textwidth]{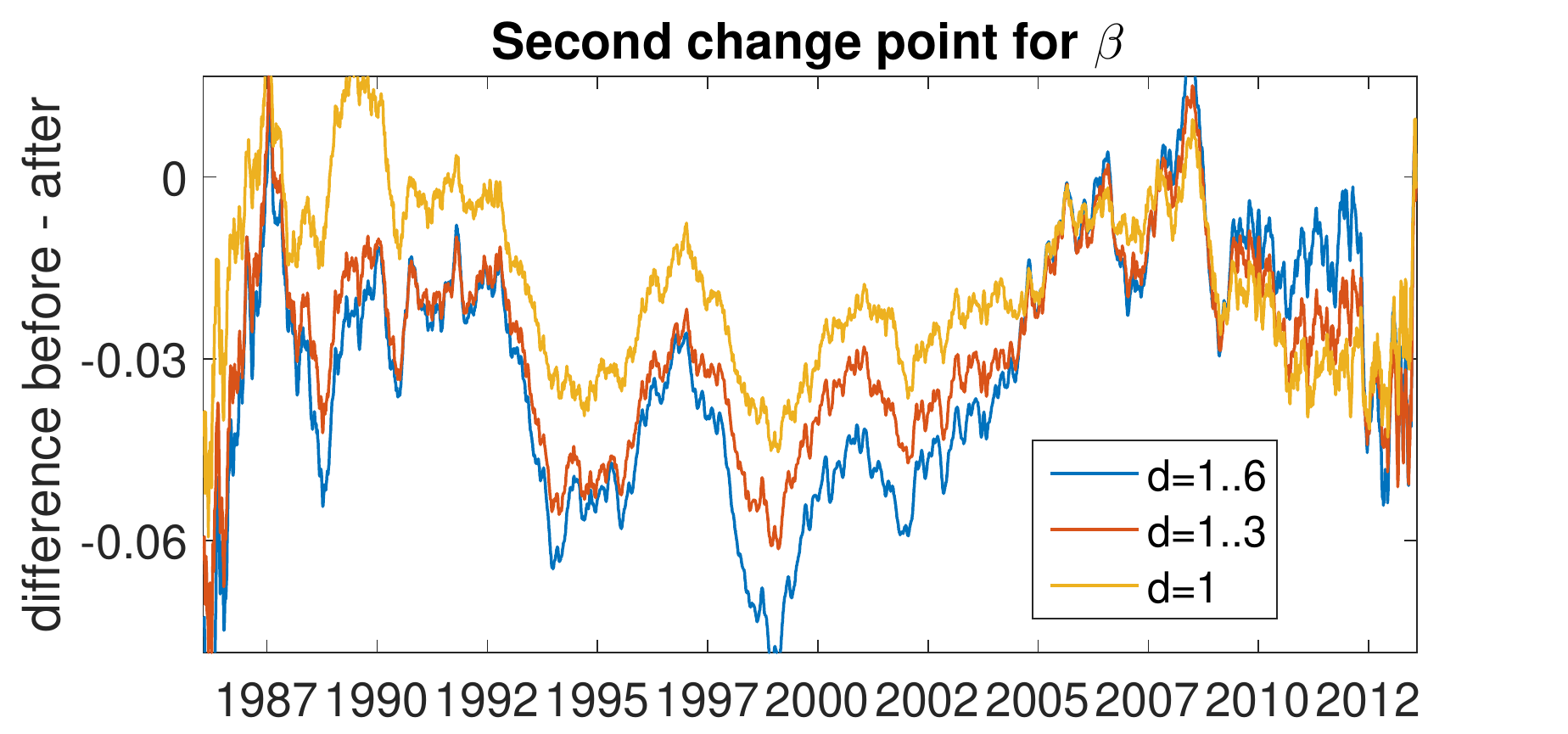} 
\includegraphics[width=.35\textwidth]{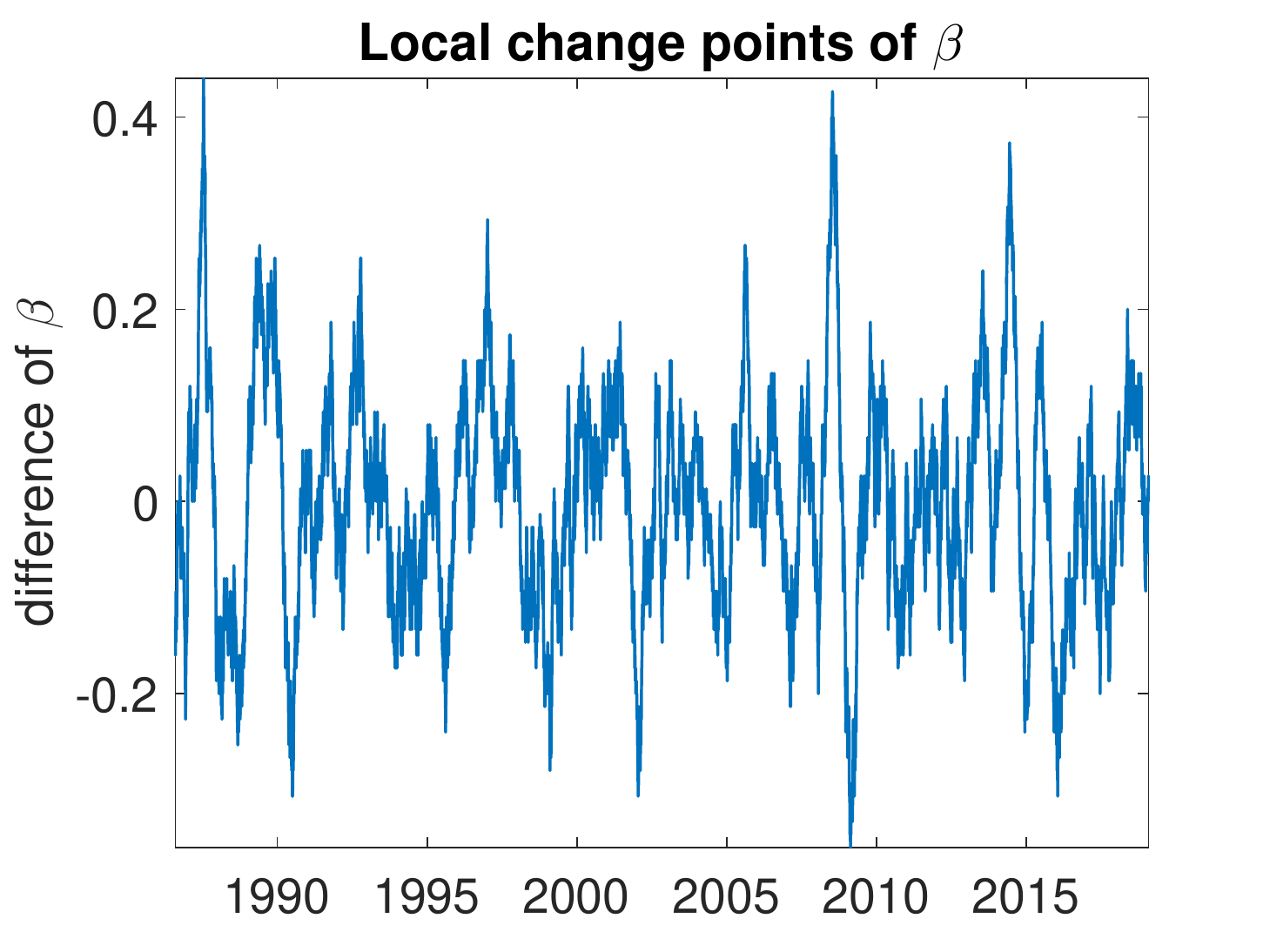}
\caption{Second change point for $\beta,$ and local change points for $\beta$ the oil prices.}\label{ordb2}
\end{figure}

As in the approach with means, we can now look for a second change point $k_2$ in the longer subinterval, in this case $[1,k_1].$ Figure \ref{ordb2} shows that the second change point is unique when we exclude margins. It is a minimum, which means that prices start to rise there, and can already be seen in Figure \ref{ordb}. Here it is comes one month later than in Figure \ref{ordch}, in early February 1999. When we look for a third change point in the interval $[k_1,k_2]$ we again get a unique solution which is already seen as maximum in Figure \ref{ordb2}. This point in July 2008 differs only by one day from our ad hoc point with the highest oil price. Thus $\beta$ produces a meaningful segmentation which is similar to, and better than, our ad hoc segmentation.

One can also search for local change points, comparing $\beta$ on the intervals $[k-m+1,k]$ and $[k+1,k+m].$ The situation is similar as for means: the local search gives a lot of maxima and minima, which depend very much on the scale parameter $m.$ Figure \ref{ordb2} shows the result for $m=150.$ Two large maxima in 2008 and 2014 are similar to those in Figure \ref{means}. Nevertheless, local search offers too many choices.

For financial data, global search with $\beta$ seems the best order method for finding change points. It provides more meaningful results than the approach with means.  

What about significance of $h(k_1)$\,?  In view of Proposition 3, we can think about random walk statistics to provide $p$-values.  Calculation of $\beta$ on intervals $[1,k]$ can be seen as a simple symmetric random walk $b_1,...,b_T$ with $T-1$ time steps. The initial value is $b_1=0,$ and the terminal value $b_T$ is the observed total number of up-steps. The probability $p_c$ that the walk will hit the lines $y=\pm c$ can be expressed by binomial coefficients.

This exciting idea has several shortcomings. First, we need the difference between relative frequencies, normalized by $c_k,$ while random walk concerns absolute values. Next, it would work only for lag 1, which from a practical viewpoint is not the best option for change point detection.in financial time series. The most serious problem, however, is that the question is wrong. It is easy to define changes in the mean or dynamics of model series. But do we really expect similar dynamical changes in a given real-world time series?

For $\beta$  as mean over lags 1 to 3, we have the maximum value $h(k_1)=0.548$ in Figure \ref{ordb}.
We checked the significance of this change point by simulation, with Brownian motion as null model. The maximum value  of $|h(k)|=c_k\cdot |\beta_k-\tilde{\beta_k}|$ was determined for $1000$ trajectories of BM. It turned out that for 422 simulations, the maximum was larger than 0.548. Almost every second realization of BM had a change point with larger value $|h(k_1)|$ than our data.

This is alright! BM is not stationary. A time series taken from BM needs segmentation, like the data. Any reasonable method should detect change points in BM. As a null model, we better take a stationary series which by definition has no change points. We first took an AR(1) model $x_t=0.99 x_{t-1}+z_t$ with i.i.d. Gaussian random numbers $z_t.$ We found that only 2 from 1000 realizations have larger $|h(k_1)|$ than our data. So for stationary series, we get significantly smaller values in the search for change points.

\begin{figure}[h]  
\centering
\includegraphics[width=.63\textwidth]{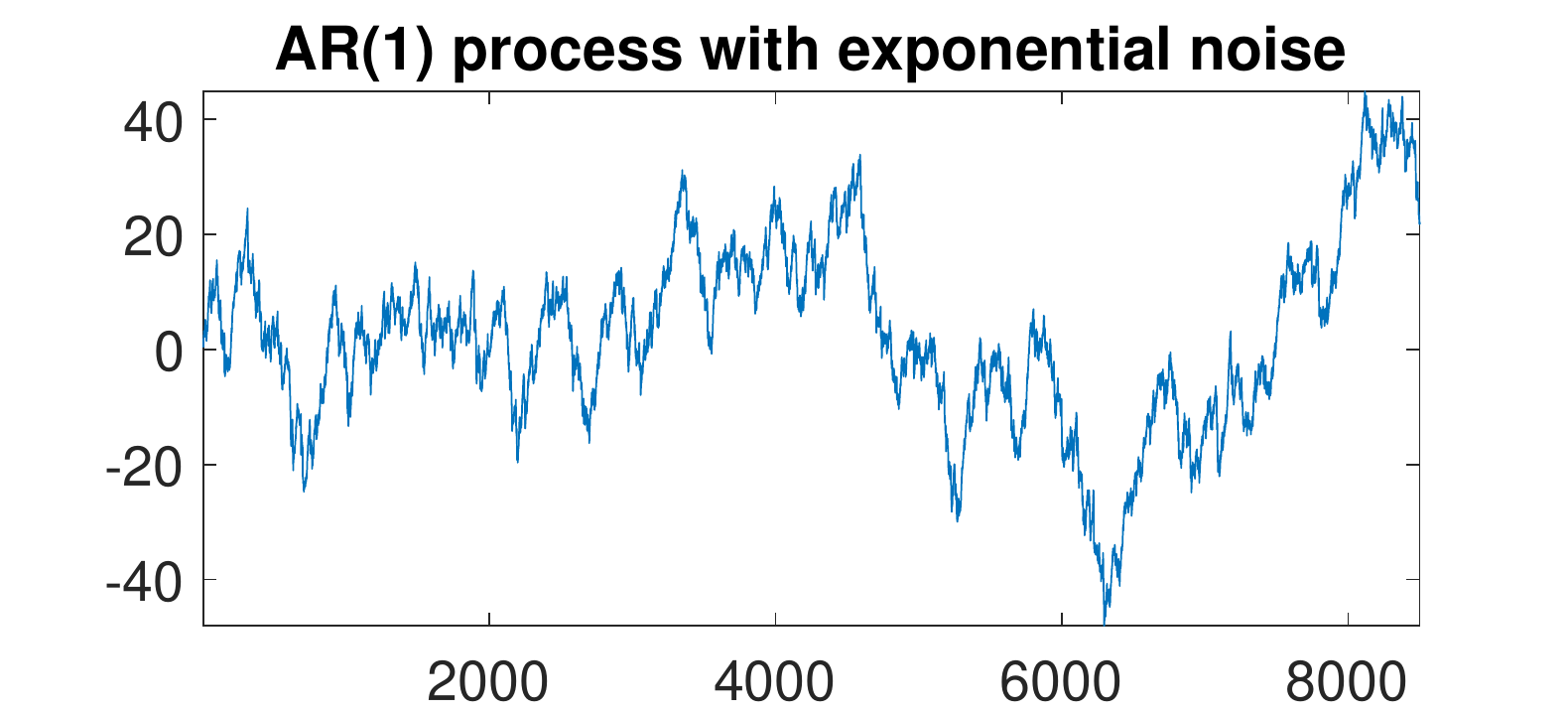} 
\caption{Sample from an AR(1) with exponential noise, with positive $\beta (d)$ for small $d.$ Although the process is stationary by definition, this cannot be seen in our figure, and segmentation may be necessary.}\label{last}
\end{figure}  

We checked this with the same AR(1) process with exponential noise, $z_t=1-e_t,$ where $e_t$ has an exponential distribution with mean 1. These random numbers have their heavy tail on the left, so the process will make downwards excursions and only slowly climb up. As a result, $\beta(1)\approx 0.25,$ and all $\beta(d)$ with $d\le 6$ are still greater 0.1. Nevertheless, only two from 1000 simulations gave a maximum larger 0.548. Then we realized that the excursions are still too short, and increased the AR(1) factor to 0.998. A typical realization with maximum $h$-value 0.32 is shown in Figure \ref{last}. For this process, 92 from 1000 simulations had maxima greater 0.548.

These experiments gave the impression that $\beta$ is a useful parameter for change point detection. It will provide large values of $h$ when we expect them. But this is exploratory research, there is still much work to do.  We hope that readers feel encouraged to perform their own studies with order patterns.


\begin{thebibliography}{10}

\bibitem{AKK}
J.~Amigo, K.~Keller, and J.~Kurths.
\newblock Recent progress in symbolic dynamics and permutation complexity. ten
  years of permutation entropy.
\newblock {\em Eur. Phys. J. Special Topics}, 222:247--257, 2013.

\bibitem{Ami}
J.~M. Amigo.
\newblock {\em Permutation complexity in dynamical systems}.
\newblock Springer Seires in Synergetics. Springer, Berlin, 2010.

\bibitem{AKU}
J.~M. Amigo, K.~Keller, and V.~A. Unakafova.
\newblock Ordinal symbolic analysis and its application to biomedical
  recordings.
\newblock {\em Phil. Trans. R. Soc. A}, 373:20140091, 2015.

\bibitem{Ba15}
C.~Bandt.
\newblock Permutation entropy and order patterns in long time series.
\newblock In I.~Rojas and H.~Pomares, editors, {\em Time Series Analysis and
  Forecasting}, Contributions to Statistics. Springer, 2015.

\bibitem{Ba18}
C.~Bandt.
\newblock Crude eeg parameter provides sleep medicine with well-defined
  continuous hypnograms.
\newblock arXiv1710.00559, 2018.

\bibitem{BP}
C.~Bandt and B.~Pompe.
\newblock Permutation entropy: a natural complexity measure for time series.
\newblock {\em Phys. Rev. Lett.}, 88:174102, 2001.

\bibitem{BS}
C.~Bandt and F.~Shiha.
\newblock Order patterns in time series.
\newblock {\em J. Time Series Analysis}, 28(5):646--665, 2007.

\bibitem{Ba17}
Christoph Bandt.
\newblock A new kind of permutation entropy used to classify sleep stages from
  invisible {EEG} microstructure.
\newblock {\em Entropy}, 19(5):197, 2017.

\bibitem{Ba19}
Christoph Bandt.
\newblock Small order patterns in big time series: a practical guide.
\newblock {\em Entropy}, 21(6):613, 2019.

\bibitem{BDS}
A.~Betken, J.~Buchsteiner, H.~Dehling, I.~Munker, A.~Schnurr, and J.~Woerner.
\newblock Ordinal patterns in long-range dependent time series.
\newblock arXiv1905.11033, 2019.

\bibitem{bien1}
I.J. Bienaym\'e.
\newblock Sur une question de probabilit\'es.
\newblock {\em Bull. Soc. Math. France}, 2:153--154, 1874.

\bibitem{bien2}
I.J. Bienaym\'e.
\newblock Application d'un th\'eor\`eme nouveau au {C}alcul des probabilit\'es.
\newblock {\em Bulletin des sciences math\'ematiques et astronomiques},
  9:219--225, 1875.

\bibitem{Bru}
A.A. Bruzzo, B.~Gesierich, M.~Santi, C.~Tassinari, N.~Birbaumer, and
  G.~Rubboli.
\newblock Permutation entropy to detect vigilance changes and preictal states
  from scalp {EEG} in epileptic patients. a preliminary study.
\newblock {\em Neurol. Sci.}, 29:3--9, 2008.

\bibitem{Chi}
B.~Chicote, U.~Irusta, R.~Alcaraz, J.~J. Rieta, E.~Aramendi, I.~Isasi,
  d.~Alonso, and K.~Ibarguren.
\newblock Application of entropy-based features to predict defibrillation
  outcome in cardiac arrest.
\newblock {\em Entropy}, 18:313, 2016.

\bibitem{EKM}
P.~Embrechts, C.~Kl\"uppelberg, and T.~Mikosch.
\newblock {\em Modelling extremal events for insurance and finance}.
\newblock Springer, 1997.

\bibitem{EM}
P.~Embrechts and M.~Maejima.
\newblock {\em Selfsimilar Processes}.
\newblock Princeton University Press, Princeton, NJ, 2002.

\bibitem{Fe}
E.~Ferlazzo and et~al.
\newblock Permutation entropy of scalp {EEG}: a tool to investigate epilepsies.
\newblock {\em Clinical Neurophysiology}, 125:13--20, 2014.

\bibitem{physio}
A.L Goldberger, L.A.N. Amaral, L.~Glass, J.M. Hausdorff, P.Ch. Ivanov, R.G.
  Mark, J.E. Mietus, G.B. Moody, C.-K. Peng, and H.E. Stanley.
\newblock Physiobank, physiotoolkit, and physionet: Components of a new
  research resource for complex physiologic signals.
\newblock {\em Circulation}, 101(63):e215--e220, 2000.
\newblock database at \url{www.physionet.org}.

\bibitem{KL}
C.-E. Kuo and S.-F. Liang.
\newblock Automatic stage scoring of single-channel sleep {EEG} based on
  multiscale permutation entropy.
\newblock {\em Biomedical Circuits and Systems Conference (BioCAS), IEEE},
  pages 448--451, 2011.

\bibitem{ManF}
Benoit~B. Mandelbrot.
\newblock {\em Fractals and Scaling in Finance}.
\newblock Springer, New York, 1997.

\bibitem{McCullough17}
M~McCullough, M~Small, HHC Iu, and T~Stemler.
\newblock Multiscale ordinal network analysis of human cardiac dynamics.
\newblock {\em Philosophical Transactions of the Royal Society A: Mathematical,
  Physical and Engineering Sciences}, 375(2096):20160292, 2017.

\bibitem{Mor}
F.~C. Morabito, D.~Labate, F.~La Foresta, A.~Bramanti, G.~Morabito, and
  I.~Palamara.
\newblock Multivariate multi-scale permutation entropy for complexity analysis
  of {A}lzheimer's disease {EEG}.
\newblock {\em Entropy}, 14:1188--1202, 2012.

\bibitem{NG}
N.~Nicolaou and J.~Georgiou.
\newblock The use of permutation entropy to characterize sleep encephalograms.
\newblock {\em Clin. EEG Neurosci.}, 39:202--209, 2012.

\bibitem{OSD}
E.~Olofsen, J.W. Sleigh, and A.~Dahan.
\newblock Permutation entropy of the electroencephalogram: a measure of
  anaesthetic drug effect.
\newblock {\em Br. J. Anaesth.}, 101:810--821, 2008.

\bibitem{ODRL}
G.~Ouyang, C.~Dang, D.A. Richards, and X.~Li.
\newblock Ordinal pattern based similarity analysis for eeg recordings.
\newblock {\em Clinical Neurophysiology}, 121:694--703, 2010.

\bibitem{Par12}
Ulrich Parlitz, Sebastian Berg, Stefan Luther, Alexander Schirdewan, J{\"u}rgen
  Kurths, and Niels Wessel.
\newblock Classifying cardiac biosignals using ordinal pattern statistics and
  symbolic dynamics.
\newblock {\em Computers in biology and medicine}, 42(3):319--327, 2012.

\bibitem{Schnurr}
Alexander Schnurr.
\newblock An ordinal pattern approach to detect and to model leverage effects
  and dependence structures between financial time series.
\newblock {\em Stat. Papers}, 55(4):919--931, 2014.

\bibitem{SD}
Alexander Schnurr and Herold Dehling.
\newblock Testing for structural breaks via ordinal pattern dependence.
\newblock {\em J. Amer. Stat. Assoc.}, 112:706--720, 2017.

\bibitem{SK11}
M.~Sinn and K.~Keller.
\newblock Estimation of ordinal pattern probabilities in {G}aussian processes
  with stationary increments.
\newblock {\em Computational Statistics and Data Analysis}, 55:1781--1790,
  2011.

\bibitem{SKC}
M.~Sinn, K.~Keller, and B.~Chen.
\newblock Segmentation and classification of time series using ordinal pattern
  distributions.
\newblock {\em Eur. Phys. J. Special Topics}, 222:587--598, 2013.

\bibitem{Te}
M.G. Terzano, L~Parrino, A~Sherieri, R~Chervin, S~Chokroverty, C~Guilleminault,
  M~Hirshkowitz, M~Mahowald, H~Moldofsky, A~Rosa, R~Thomas, and A~Walters.
\newblock Atlas, rules, and recording techniques for the scoring of cyclic
  alternating pattern (cap) in human sleep.
\newblock {\em Sleep Med.}, 2(6):537--553, 2001.

\bibitem{UK18}
Anton~M. Unakafov and Karsten Keller.
\newblock Change-point detection using the conditional entropy of ordinal
  patterns.
\newblock {\em Entropy}, 20:709, 2018.

\bibitem{ZZRP}
M.~Zanin, L.~Zunino, O.A. Rosso, and D.~Papo.
\newblock Permutation entropy and its main biomedical and econophysics
  applications: a review.
\newblock {\em Entropy}, 14:1553--1577, 2012.

\bibitem{ZB12}
L.~Zunino, A.F. Bariviera, M.B. Guercio, L.B. Martinez, and O.A. Rosso.
\newblock On the efficiency of sovereign bond markets.
\newblock {\em Physica A}, 391:4342--4349, 2012.

\bibitem{ZB16}
L.~Zunino, A.F. Bariviera, M.B. Guercio, L.B. Martinez, and O.A. Rosso.
\newblock Monitoring the informational efficiency of {E}uropean corporate bond
  markets with dynamical permutation min-entropy.
\newblock {\em Physica A}, 456:1--9, 2016.

\bibitem{ZZTP}
L.~Zunino, M.~Zanin, B.M. Tabak, D.G. P\'erez, and O.A. Rosso.
\newblock Forbidden patterns, permutation entropy and stock market
  inefficiency.
\newblock {\em Physica A}, 388:28543--2864, 2009.

\end{thebibliography}
\end{document}